\documentclass{aa}    
\usepackage{txfonts}
\usepackage{longtable}  
\usepackage{natbib}
\usepackage{graphicx}

\bibpunct[, ]{(}{)}{,}{a}{}{,}

\usepackage{color}

\newcommand{\LyudmilasStar}{HE~2327$-$5642}        
\newcommand{\LyudaStar}{HE~2252$-$4225} 
\newcommand{\tefft}{$T_{\mbox{\scriptsize eff}}$}  
\newcommand{\teffm}{T_{\mbox{\scriptsize eff}}}    

\newcommand{\eps}[1]{\log\varepsilon_{\rm #1}}
\newcommand{\kms}{km\,s$^{-1}$}

\newcommand{\kH}{$S_{\!\!\rm H}$}    
\newcommand{\Eexc}{$E_{\rm exc}$}
\newcommand{\eu}[5]{\mbox{$#1\,^#2{\rm #3}^{#4}_{\rm #5}$}}

\begin{document}

\title{The Hamburg/ESO R-process Enhanced Star survey (HERES) \thanks{Based on
    observations collected at the European Southern Observatory, Paranal,
    Chile (Proposal numbers 170.D-0010, and 280.D-5011).} \thanks{Table 3 is only available in electronic form
at the CDS via anonymous ftp to cdsarc.u-strasbg.fr (130.79.128.5)
or via http://cdsweb.u-strasbg.fr/cgi-bin/qcat?J/A+A/}}
\subtitle{X. HE~2252$-$4225, one more r-process enhanced and actinide-boost halo star}

\author{
  L. Mashonkina\inst{1,2} \and
  N. Christlieb\inst{3} \and
  K. Eriksson\inst{4} 
}

\offprints{L. Mashonkina; \email{lima@inasan.ru}}
\institute{
     Universit\"ats-Sternwarte M\"unchen, Scheinerstr. 1, D-81679 M\"unchen, 
     Germany \\ \email{lyuda@usm.lmu.de}
\and Institute of Astronomy, Russian Academy of Sciences, RU-119017 Moscow, 
     Russia \\ \email{lima@inasan.ru}
\and Zentrum f\"ur Astronomie der Universit\"at Heidelberg, Landessternwarte,
     K\"onigstuhl 12, D-69117 Heidelberg, Germany \\
     \email{N.Christlieb@lsw.uni-heidelberg.de}
\and Department of Astronomy and Space Physics, Uppsala University, Box 515, 
     75120 Uppsala, Sweden
}

\date{Received  / Accepted }

\abstract{Studies of the $r$-process enhanced stars are important for understanding the nature and origin of the $r$-process better.}
{We present a detailed abundance analysis of a very metal-poor giant star discovered in the HERES project,
  HE~2252--4225, which exhibits overabundances of the $r$-process elements with [$r$/Fe] = +0.80.}
{We determined the 
  stellar atmosphere parameters, $\teffm = 4710$~K, log~g = 1.65, and [Fe/H] $= -2.63$, and chemical abundances by analysing the 
  high-quality VLT/UVES spectra. The surface gravity was calculated from the
  non-local thermodynamic equilibrium (NLTE) ionisation balance between
  \ion{Fe}{i} and \ion{Fe}{ii}.}
{Accurate abundances for a total of 38 elements, including 22 neutron-capture elements
  beyond Sr and up to Th, were determined in {\LyudaStar}. For every chemical species, the dispersion in the single line measurements around the mean does not exceed 0.12~dex. This object is deficient in carbon, as expected for a giant star with $\teffm < 4800$~K. The stellar Na--Zn abundances are well fitted by the yields of a single supernova of 14.4~$M_\odot$.
For the neutron-capture elements in the Sr--Ru, Ba--Yb, and Os--Ir regions, the 
 abundance pattern of {\LyudaStar} is in excellent agreement with
  the average abundance pattern of the strongly $r$-process enhanced
 stars CS\,22892-052, CS\,31082-001, HE\,1219-0312, and HE\,1523-091. This suggests a common origin of the first, second, and third $r$-process peak elements in {\LyudaStar} in the classical $r$-process. We tested the solar $r$-process pattern based on the most recent $s$-process calculations of Bisterzo, Travaglio, Gallino, Wiescher, and K{\"a}ppeler and found that elements in the range from
  Ba to Ir match it very well. No firm
  conclusion can be drawn about the relationship between the first
  neutron-capture peak elements, Sr to Ru, in {\LyudaStar} and the solar
  $r$-process, owing to the uncertainty in the solar $r$-process.
The investigated star has an anomalously high Th/Eu abundance ratio, so that radioactive dating  results in a stellar age of $\tau = 1.5\pm1.5$~Gyr that is not expected for a very metal-poor halo star.}
{}


\titlerunning{HE~2252$-$4225, one more r-process enhanced and actinide boost star}
\authorrunning{Mashonkina et al.}

\maketitle
%
\section{Introduction}\label{Sect:intro}

{\LyudaStar} was identified as a candidate metal-poor
star in the Hamburg/ESO Survey \citep[HES; see][for details of the candidate selection procedures]{Christliebetal:2008a}. Moderate-resolution
($\Delta\lambda = 2$\,{\AA}) spectroscopy obtained at the Siding Spring
Observatory (SSO) 2.3\,m-telescope with the Double Beam Spectrograph (DBS)
confirmed its metal-poor nature. Therefore, it was included in the target list
of the Hamburg/ESO R-process-Enhanced Star
survey (HERES). A detailed description of the project and its aims can
be found in \citet[][hereafter Paper~I]{HERESI}, and the methods of automated
abundance analysis of high-resolution ``snapshot'' spectra are described
in \citet[][hereafter Paper~II]{HERESII}.  ``Snapshot'' spectra with a spectral resolution $R\sim 20\,000$ and a signal-to-noise ratio of $S/N\sim 50$ per pixel at 4100\,{\AA} were used to show that
{\LyudaStar} is a very metal-poor (VMP) star, with the iron abundance [Fe/H]\footnote{In the classical notation, where [X/Y] = $\log(N_{\rm X}/N_{\rm Y})_{star} - \log(N_{\rm X}/N_{\rm Y})_{Sun}$.} $= -2.83$, and it exhibits overabundances of the heavy elements produced via rapid neutron captures ($r$-process), with [Eu/Fe] = $+0.99$ and $\mathrm{[Ba/Fe]} = +0.45$ (Paper\,II).  

The astrophysical site(s) of the $r$-process is still unclear. Studies of the strongly $r$-process enhanced and VMP stars can provide an empirical information about the origin of the heavy elements beyond the iron group in the early Galaxy.
\citet{HERESI} suggest referring to metal-poor stars having [Eu/Fe] = +0.3 to +1.0 
and $\mathrm{[Ba/Eu]} < 0$ as r-I stars. Stars with similar [Ba/Eu] abundance ratio, but higher $\mathrm{[Eu/Fe]} > +1$, belong to the group of r-II stars.
Among a dozen discovered r-II stars, the six stars, CS\,22892-052 \citep{Sneden1994}, CS\,31082-001 \citep{2001Natur.409..691C}, CS\,29497-004 (Paper~I), HE\,1219-0312 (Paper~II), HE\,1523-091 \citep{he1523}, and SDSS\,J2357-0052 \citep{Aoki2010}, are extremely enhanced in the $r$-process elements, with [Eu/Fe] $\ge 1.5$. Their heavy-element abundances are expected to be dominated by the influence of a single event, or at most very few $r$-process nucleosynthesis events. 

It was established that r-II and r-I stars have a very similar heavy-element abundance pattern in the Ba--Hf region that, in turn, is consistent with a scaled solar system (SS) $r$-process abundance distribution \citep[see for example][]{Sneden2008,Cowan2002,Ivans2006,hd29907_2011,2014arXiv1404.0234S}.
This suggests a universal production ratio of the second $r$-process peak elements during the Galaxy history. A different case is the light trans-iron elements. Observations of MP halo stars appear to infer a
distinct production mechanism for Sr--Zr and heavy
elements beyond Ba in the early Galaxy \citep{Aoki2005,Francois2007,Mashonkina2007}. \citet{HE2327} found a clear distinction in
  Sr/Eu abundance ratios between the r-II and r-I stars, namely, the r-II stars contain a low
  Sr/Eu abundance ratio at [Sr/Eu] = $-0.92\pm0.13$, while the r-I stars have 0.36\,dex higher Sr/Eu values.

In this paper, we investigate whether the heavy-element abundance pattern of {\LyudaStar} matches that of the extremely neutron-capture-rich r-II  stars. What is the behaviour of the first $r$-process peak elements from Sr to Pd in {\LyudaStar}? Can a nucleo-chronometric age of {\LyudaStar} be
estimated from measurement of the radioactive element thorium? How can an abundance analysis of {\LyudaStar} improve our knowledge of the $r$-process? 
To answer these questions, we continue our series of papers on the HERES project and aim to revise stellar parameters and to perform detailed abundance analysis of {\LyudaStar}, based on the high-quality VLT/UVES spectra and refined theoretical methods of line-formation modelling.

This paper is structured as follows. After presenting the observations in 
Sect.~\ref{Sect:Observations}, we describe our determination of stellar parameters and abundance analysis of
{\LyudaStar} in Sects.~\ref{Sect:stellarParameters} and \ref{Sect:AbundanceAnalysis}. Section~\ref{Sect:ssr} analyses the heavy element abundance pattern of the investigated star. An anomalously high Th abundance of {\LyudaStar} is discussed in Sect.~\ref{Sect:Th}. Our conclusions are presented in Sect.~\ref{Sect:DiscussionConclusions}.

\section{Observations}\label{Sect:Observations}

For the convenience of the reader, we list the coordinates and photometry of {\LyudaStar} in Table~\ref{Tab:CoordsPhotometry}. The photometry was taken from \citet{Beersetal:2007}. High-quality spectra of
this star was acquired during May--September 2005 with the VLT and UVES in
dichroic mode. The BLUE390$+$RED580 (9\,h total integration time), and
BLUE437$+$RED860 (10\,h) standard settings were employed to ensure a wide
wavelength coverage. The slit width of $0.8''$ in both arms yielded a
resolving power of $R=50\,000$. A $1\times 1$ pixel binning ensured proper
sampling of the spectra. The observations are summarised in
Table~\ref{Tab:Observations}.

The pipeline-reduced spectra were shifted to the stellar rest frame and then co-added
in an iterative procedure in which we identified pixels in the individual spectra affected
by cosmic ray hits that had not been fully removed during the data reduction, or those affected by
CCD defects or other artefacts. These pixels were flagged
and ignored in the final iteration of the co-addition. Both sets of co-added blue
spectra have $S/N$ of at least 50 per pixel at
$\lambda > 3800$\,{\AA}.  At the shortest wavelengths, the $S/N$ of the
BLUE390 and BLUE437 is $10$ (at 3400\,{\AA}) and 70 (at 3756\,{\AA}),
respectively. The red arm spectra have $S/N > 100$ per pixel in most of the
covered spectral range.

\begin{table}[htbp]
 \centering
 \caption{\label{Tab:CoordsPhotometry} Coordinates and photometry of 
   {\LyudaStar}. }
  \begin{tabular}{lr}
   \hline\hline
   R.A.(2000.0) & 22:54:58.6 \\
   dec.(2000.0) & $-$42:09:19 \\
   $V$          & $14.878 \pm 0.003$\\
   $B-V$        & $ 0.822\pm 0.005$\\
   $V-R$        & $ 0.499 \pm 0.005$\\
   $V-I$        & $ 1.023\pm 0.006$\\
   \hline
  \end{tabular}
\end{table}

\begin{table}[htbp]
 \centering
 \caption{\label{Tab:Observations} VLT/UVES observations of
   {\LyudaStar}. }
  \begin{tabular}{ccrr}\hline\hline
   Setting  & $\lambda^1$  & \multicolumn{1}{c}{$t_{\mbox{\scriptsize exp}}$} & \multicolumn{1}{c}{$S/N^2$} \\\hline
   BLUE390  & $3400$--$4515$\,{\AA} & $ 8.9$\,h &  $10$--$65$ \\ 
   BLUE437  & $3765$--$4985$\,{\AA} & $10.0$\,h &  $70$--$120$\\ 
   REDL580  & $4795$--$5760$\,{\AA} & $ 8.9$\,h & $100$--$120$\\ 
   REDU580  & $5845$--$6810$\,{\AA} & $ 8.9$\,h &  $60$--$120$\\ 
   REDL860  & $8400$--$8526$\,{\AA} & $10.0$\,h & $100$--$180$ \\\hline 
\multicolumn{4}{l}{Notes. $ ^1$ \ $\lambda$ refers to rest frame wavelengths,} \\
\multicolumn{4}{l}{$ ^2$ \ $S/N$ refers to the signal-to-noise ratio per pixel.} \\
\end{tabular}
\end{table}

\section{Stellar atmosphere parameters}\label{Sect:stellarParameters}

\begin{figure} 
  \resizebox{88mm}{!}{\includegraphics{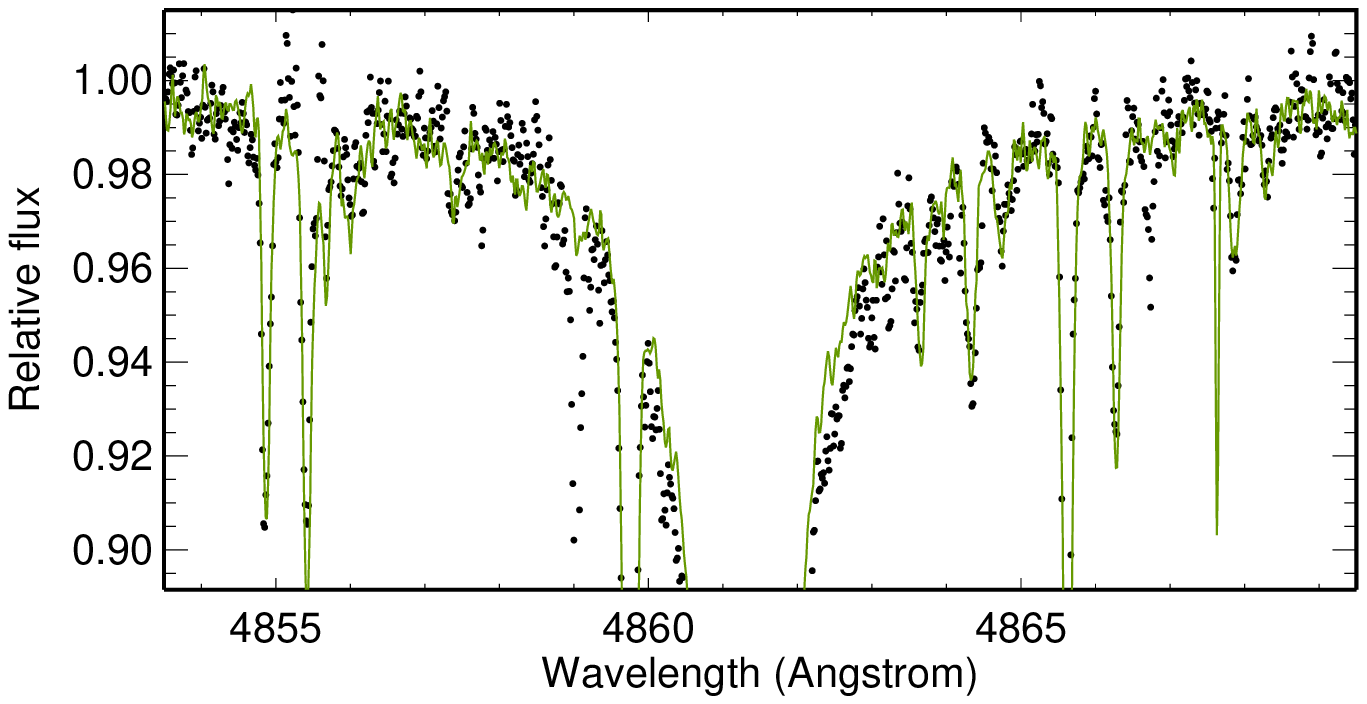}}

  \vspace{-5mm}
  \resizebox{88mm}{!}{\includegraphics{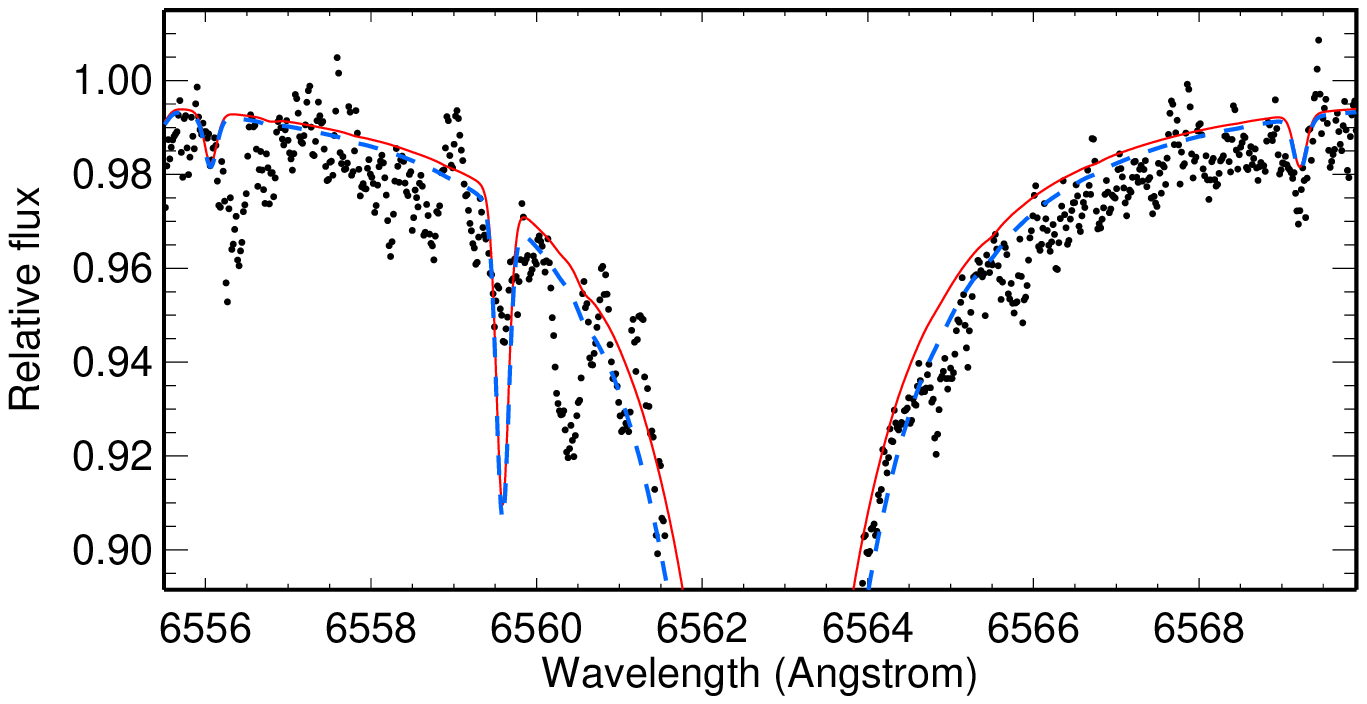}}

  \vspace{-5mm}
  \resizebox{88mm}{!}{\includegraphics{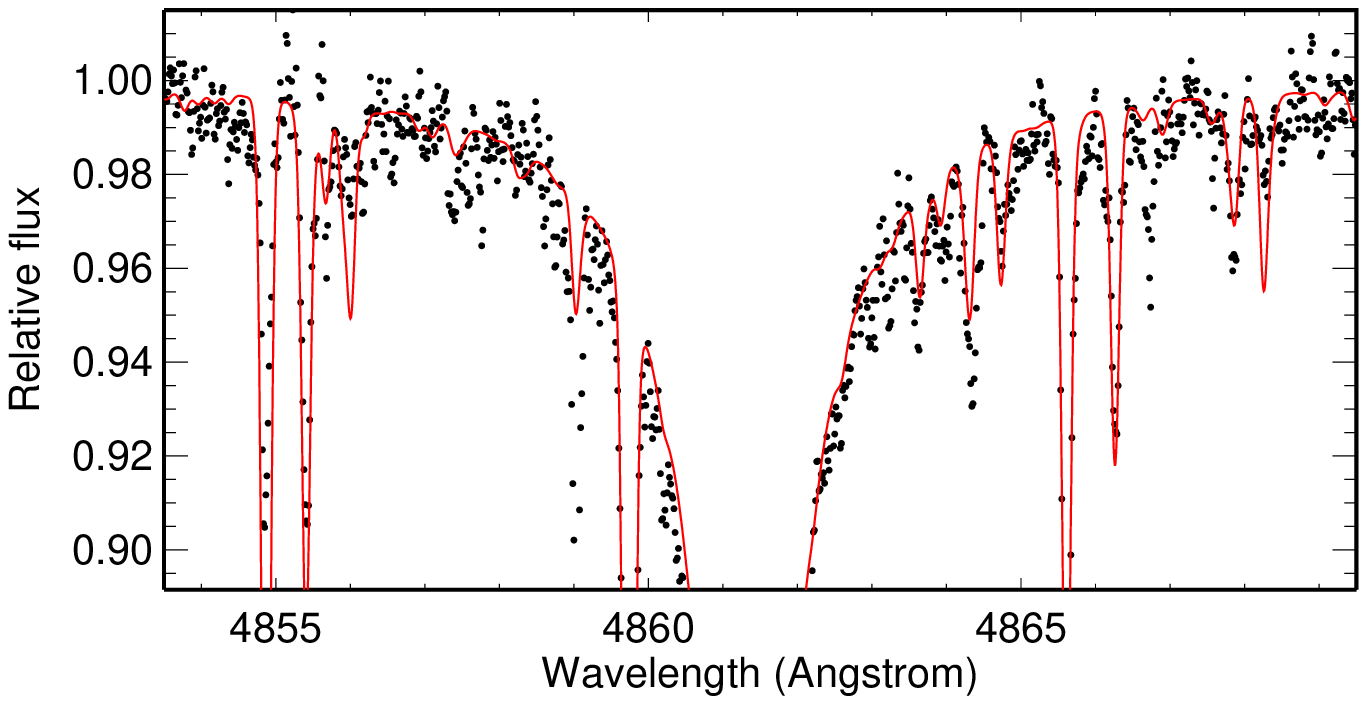}}

  \caption{\label{Fig:balmer} Top panel: observed profiles of H$\beta$ in {\LyudaStar} (bold dots) and HD\,122563 (continuous curve). 
Middle and bottom panels: synthetic NLTE (continuous curve) and LTE (dashed curve) flux profile of H$\alpha$ and H$\beta$ computed for $\teffm = 4710$\,K compared to the observed spectrum of {\LyudaStar} (bold dots). In all calculations, $\log g = 1.65$, $\mathrm{[Fe/H]} = -2.66$, and $\xi = 1.7$\,\kms. }
\end{figure}

 In Paper~II, an effective temperature of $\teffm = 4708 \pm 100$\,K was determined from
photometry when adopting the reddening derived from the maps of
\citet{1998ApJ...500..525S}. A subsequent analysis of the snapshot spectrum inferred that {\LyudaStar} is a VMP giant with the surface gravity $\log g = 1.53\pm0.24$, [Fe/H] = $-2.83\pm 0.12$ and the microturbulence velocity $\xi_t = 1.9\pm0.1$~\kms. 
These parameters appear to be close to those of HD\,122563, which is one of the best observed halo stars, with well-determined \tefft\ and log~g. Recent measurements of the angular diameter of HD\,122563 resulted in $\teffm$ = 4600$\pm$41~K \citep{2012A&A...545A..17C}. Employing the same \tefft, 
\citet{mash_fe} calculated log~g = 1.60$\pm$0.07 from the {\sc HIPPARCOS} parallax and [Fe/H] = $-2.56$ from the NLTE analysis of the Fe lines. 
 We found that {\LyudaStar} and HD\,122563 have very similar line profiles for H$\alpha$ and H$\beta$ (Fig.\,\ref{Fig:balmer}, top panel displays only H$\beta$). This confirms that {\LyudaStar} is indeed cool, and its effective temperature may be only slightly higher than that of HD\,122563. The high-quality spectrum of HD\,122563 ($R \simeq 80\,000$ and S/N $> 200$) was taken from the {\sc ESO UVESPOP} survey \citep{2003Msngr.114...10B}.  

In this study, we check $\teffm = 4710$\,K through a profile
analysis of H$\alpha$ and H$\beta$ and revise the surface gravity, iron abundance, and microturbulence velocity using lines of \ion{Fe}{i} and \ion{Fe}{ii}. H$\gamma$ was not used due to heavy blending of the line wings. Our calculations are based on non-local thermodynamic equilibrium (NLTE) line formation for \ion{H}{i} and \ion{Fe}{i-ii}, employing the methods described by \citet{Mashonkina2008} and \citet{mash_fe}, respectively. The coupled radiative transfer and statistical equilibrium (SE) equations were solved with a revised version of the DETAIL code \citep{detail}. The update was presented by \citet{mash_fe}. The obtained departure coefficients were then used by the code SIU \citep{Reetz} to calculate the synthetic line profiles. 

When determining the stellar parameters of {\LyudaStar}, we used the MARCS model structures \citep{Gustafssonetal:2008}\footnote{\tt http://marcs.astro.uu.se}, which were interpolated for given \tefft, log~g, and [Fe/H] using a FORTRAN-based routine written by Thomas Masseron\footnote{{\tt http://marcs.astro.uu.se/software.php}}. 

\subsection{Effective temperature}

 The Balmer lines were computed for a small grid of model atmospheres with common $\teffm = 4710$\,K, but varying gravity, with log~g = 1.2, 1.53, 1.65, and 1.75, and metallicity, with [M/H] = $-2.5$, $-2.66$, $-2.83$, and $-3.0$. The theoretical profiles were obtained by convolving the profiles resulting from the thermal, natural, and Stark
broadening \citep{1973ApJS...25...37V}, as well as self-broadening \citep{BPO}. In the SE calculations, inelastic collisions with hydrogen atoms were accounted for using the
classical Drawin rates \citep{Drawin1968,Drawin1969} scaled by a factor of \kH\,= 2, as recommended by \citet{Mashonkina2008}. 
It was found that a variation in log~g and [M/H] within 0.24~dex and 0.12~dex, respectively, produces minor effect on the H$\alpha$ and H$\beta$ profiles, such that the difference in derived {\tefft} does not exceed 30\,K. 
Figure~\ref{Fig:balmer} displays observed profiles of the two Balmer lines in {\LyudaStar} compared with the theoretical NLTE and LTE (only for H$\alpha$) ones calculated for $\log g = 1.65$ and $\mathrm{[Fe/H]} = -2.66$. The departures from LTE are weak for the H$\beta$ profile beyond
the core, such that the difference between {\tefft} derived for this line, assuming either NLTE or LTE, does not exceed 30\,K. For both lines, the observed wings are satisfactorily reproduced by the model atmosphere with $\teffm = 4710$\,K, and this value was adopted as a final effective temperature of {\LyudaStar}. 

\begin{figure} 
  \resizebox{88mm}{!}{\includegraphics{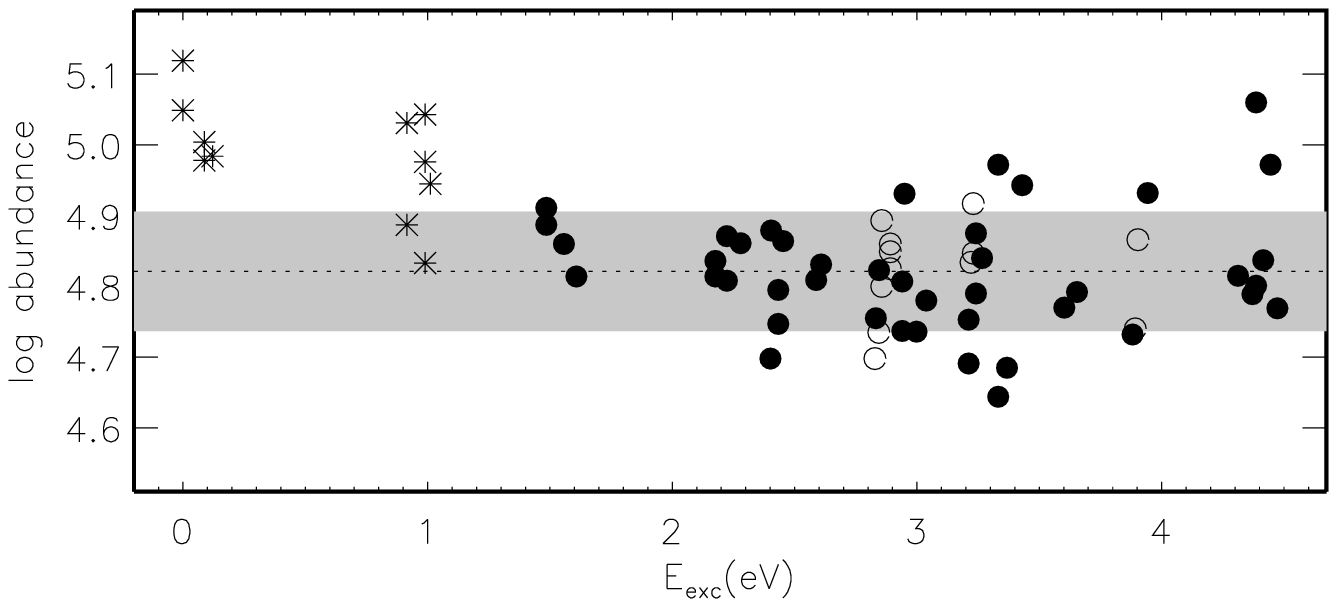}}

  \vspace{-5mm}
  \resizebox{88mm}{!}{\includegraphics{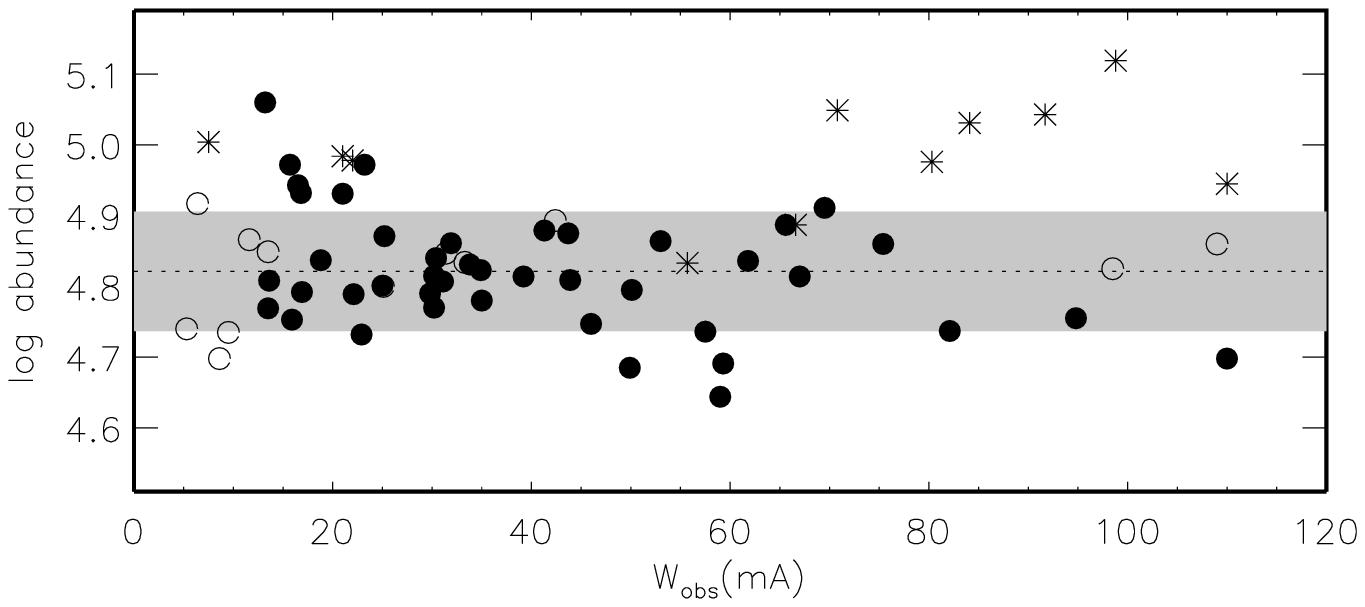}}

  \caption{\label{Fig:fe_w_exc}  Trends of NLTE abundances with excitation potential and equivalent width, as determined from individual \ion{Fe}{i}
    (filled circles for \Eexc\ $\ge$ 1.2~eV and asterisks for \Eexc\ $<$ 1.2~eV) and \ion{Fe}{ii} (open circles) lines, using our adopted
    stellar parameters. The dotted line indicates the mean Fe abundance from
    two ionisation stages and the shaded grey area its statistical error.}
\end{figure}

\subsection{Surface gravity, iron abundance, and microturbulence velocity}

The lines employed to determine log~g, [Fe/H], and $\xi_t$ were mainly selected from the linelist produced in our previous HERES paper \citep[][hereafter, Paper~V]{HE2327}. They are listed in Table~\ref{linelist} (online material), along with transition information, references for the adopted $gf$-values, and the final element abundances. The LTE and NLTE abundances were calculated for the four model atmospheres, with common $\teffm = 4710$\,K and [M/H] = $-2.83$, but various log~g = 1.20, 1.53, 1.65, and 1.75 and $\xi_t$ = 1.7, 1.9, and 2.1\,\kms. In the atmospheres of MP stars, the main source of uncertainties in NLTE results is poorly known inelastic collisions with \ion{H}{i} atoms. 
 Therefore, the two sets of NLTE abundances corresponding to \kH\ = 0.1 and \kH\ = 1 were computed for each model. Figure~\ref{Fig:fe_w_exc} displays the NLTE(\kH\ = 0.1) abundances from individual lines of \ion{Fe}{i} and \ion{Fe}{ii} in the model $\teffm$/log~g/[M/H]/$\xi_t$ = 4710/1.65/-2.66/1.7. 

We found that the low-excitation lines (\Eexc\ $<$ 1.2~eV) of \ion{Fe}{i} give systematically higher abundances compared with the mean from the remaining lines, independent of either LTE or NLTE and independent of the used $\xi_t$ and log~g. For example, in the model 4710/1.65/-2.66/1.7, the abundance difference between the \Eexc\ $<$ 1.2~eV and \Eexc\ $>$ 1.2~eV lines of \ion{Fe}{i} amounts to 0.19~dex in NLTE(\kH\ = 0.1) and 0.21~dex in LTE. This can be caused by ignoring hydrodynamic phenomena (3D effects) in the atmosphere of {\LyudaStar}. \citet{Collet2007,Hayek2011}, and \citet{2013A&A...559A.102D} predicted that the (3D-1D) abundance corrections are negative for lines of \ion{Fe}{i} in the [M/H] $= -2$ and $-3$ models of VMP cool giants and the magnitude of the correction depends strongly on excitation energy of the lower level. For example, in the 4858/2.2/$-3$ model, (3D-1D) = $-0.8$~dex and $-0.05$~dex for the \Eexc\ = 0 and 4~eV lines, respectively \citep{Hayek2011}. We do not see such a large discrepancy between the low- and high-excitation lines of \ion{Fe}{i} in {\LyudaStar}. Nevertheless, we excluded lines with \Eexc\ $<$ 1.2~eV, when calculating the mean abundance and analysing the abundance trend with the line strength for \ion{Fe}{i}.

It was found that a variation of 0.2~\kms\ in $\xi_t$ led to a similar change of 0.03~dex in the derived mean abundances from \ion{Fe}{i} and \ion{Fe}{ii} lines, therefore, the gravity determination did not depend on the adopted value of $\xi_t$. However, including the results for $\xi_t = 2.1$\,\kms\ produced a steep trend with observed equivalent widths, $W_{\rm obs}$, for the 
abundances found from individual \ion{Fe}{i} and \ion{Fe}{ii} lines,
independent of the adopted values of log~g.
The least slope of the $\eps{Fe}$ - $W_{\rm obs}$ plot was obtained with $\xi_t = 1.7$\,\kms.

Consistent iron abundances for the two ionisation stages were obtained with log~g = 1.2 in LTE, log~g = 1.35 in NLTE with \kH\ = 1, and log~g = 1.65 in NLTE with \kH\ = 0.1. The discrepancy in LTE gravity between this study and Paper~II can be explained by using different sources of $gf$-values for \ion{Fe}{ii} lines and a different list of the Fe lines in the two studies. The present results are based on recent $gf$-values of \citet{MB09}. When choosing between \kH\ = 1 and 0.1, 
 we relied on an analysis of iron lines in HD\,122563, where applying \kH\ = 0.1 led to
consistent NLTE abundances for \ion{Fe}{i} and \ion{Fe}{ii} for the gravity calculated from the {\sc HIPPARCOS} parallax \citep{mash_fe}. Thus, log~g = 1.65 was adopted as a final surface gravity of {\LyudaStar}. 
The uncertainty in log~g was estimated as 0.15~dex based on the statistical errors of the mean abundances for \ion{Fe}{i}, $\eps{FeI}$ = 4.79$\pm$0.09, and \ion{Fe}{ii}, $\eps{FeII}$ = 4.79$\pm$0.07. We refer to abundances on the usual scale, where $\eps{H} = 12$. Hereafter,  
the statistical error is the dispersion in the single line measurements around the mean: $\sigma_{\eps{}} = \sqrt{\Sigma(\overline{x}-x_i)^2/(N_{lines}-1)}$. 

We did not use the ionisation equilibrium between \ion{Ca}{i} and \ion{Ca}{ii} to constrain the gravity because the observed \ion{Ca}{ii} 3933\,\AA\ and 8498\,\AA\ lines are very strong, with $W_{\rm obs}$ = 5\,848~m\AA\ and 509~m\AA, respectively, and they cannot be accurately measured due to the uncertainty in continuum normalisation. Furthermore, \ion{Ca}{ii} 8498\,\AA\ is insensitive to surface gravity variation. 
The difference in abundance from \ion{Ca}{ii} 8498\,\AA\ amounts to $-0.03$~dex between log~g = 1.20 and 1.53, and it is smaller than 0.01~dex between log~g = 1.53, 1.65, and 1.75.

\begin{table} 
  \renewcommand\thetable{4}
 \centering
 \caption{\label{Tab:StellarParameter} Determined stellar parameters of 
   {\LyudaStar}.}
  \begin{tabular}{ccl}
   \hline\hline
   Parameter & Value & Uncertainty \\
   \hline
   {\tefft} & 4710 K & $\pm100$K \\
   $\log g$ & 1.65   & $\pm0.15$\\
   $\mathrm{[Fe/H]}$ & $-2.63$ & $\pm0.08$\\
   $ \xi_t$        & 1.7\,\kms & $ \pm0.1$\kms \\
   \hline
  \end{tabular}
\end{table}

The obtained parameters $\teffm = 4710$\,K, log~g = 1.65, [Fe/H] = $-2.66$, and $ \xi$ = 1.7\,\kms\ were employed to compute the final model atmosphere with the code MARCS. Enhancements of the $\alpha$-elements O, Mg, Si, and Ca were adopted to be typical of galactic halo stars, with [$\alpha$/Fe] = 0.4.
 We found that moving from the interpolated to an explicitly tailored model atmosphere does not require any revision in derived surface gravity and microturbulence velocity, and the mean NLTE abundances from lines of \ion{Fe}{i} and \ion{Fe}{ii} increase by a similar and small amount of 0.03~dex. With the new model atmosphere, they equal $\eps{FeI}$ = 4.82$\pm$0.08 and $\eps{FeII}$ = 4.82$\pm$0.07. We also checked lines of CH, \ion{Ca}{i}, and \ion{Ba}{ii}. For the atomic lines, the difference nowhere exceeds 0.03~dex. However, the use of the interpolated model leads to a 0.08~dex lower C abundance from the CH molecular lines. 
The adopted stellar parameters of {\LyudaStar} are given in Table~\ref{Tab:StellarParameter}. The investigated star is, most probably, a distant object, with a spectroscopic distance of 12.6 kpc, as estimated assuming a stellar mass of 0.8 solar mass.

\section{Abundance analysis}\label{Sect:AbundanceAnalysis}

Our determinations of the elemental abundances are
based on line-profile and equivalent-width analyses, using the codes SIU \citep{Reetz} and WIDTH9\footnote{\tt http://kurucz.harvard.edu/programs/WIDTH/} \citep{2005MSAIS...8...14K}, respectively. The SIU and WIDTH9 codes both treat continuum scattering correctly; i.e.,
scattering is taken into account not only in the absorption coefficient, but
also in the source function. 

The lines used in the abundance analysis are listed in Table~\ref{linelist} (online material), along with the transition information and references to the adopted $gf$-values. They 
were mostly selected from the list produced in Paper~V. The data from Paper~V were replaced with $gf$-values from recent laboratory measurements for \ion{Ti}{i} \citep{Lawler2013_ti1}, \ion{Ti}{ii} \citep{2013_gf_ti2}, \ion{Mn}{i} \citep{DLSSC}, and \ion{Zn}{i} \citep{2012ApJ...750...76R}, where available.
Hyper-fine splitting (HFS) and/or isotopic splitting (IS) structure were accounted for properly for the lines of chemical elements that are represented by either a single
isotope with an odd number of nucleons or multiple isotopes. Table~\ref{linelist} (online material) provides notes that indicate whether HFS/IS was considered in a given feature, and references to the HFS/IS data used. For \ion{Mn}{i} levels, we used magnetic dipole constants $A$ from Table~1 in \citet{2003A&A...404.1153L}. The method of calculations and adopted isotope abundances were described in detail in Paper~V. 
 We ignored any lines with equivalent widths larger than 100\,m{\AA}. Exceptions were the elements, such as strontium, for which only strong lines can be detected in {\LyudaStar}.

\begin{figure*}  
  \resizebox{88mm}{!}{\includegraphics{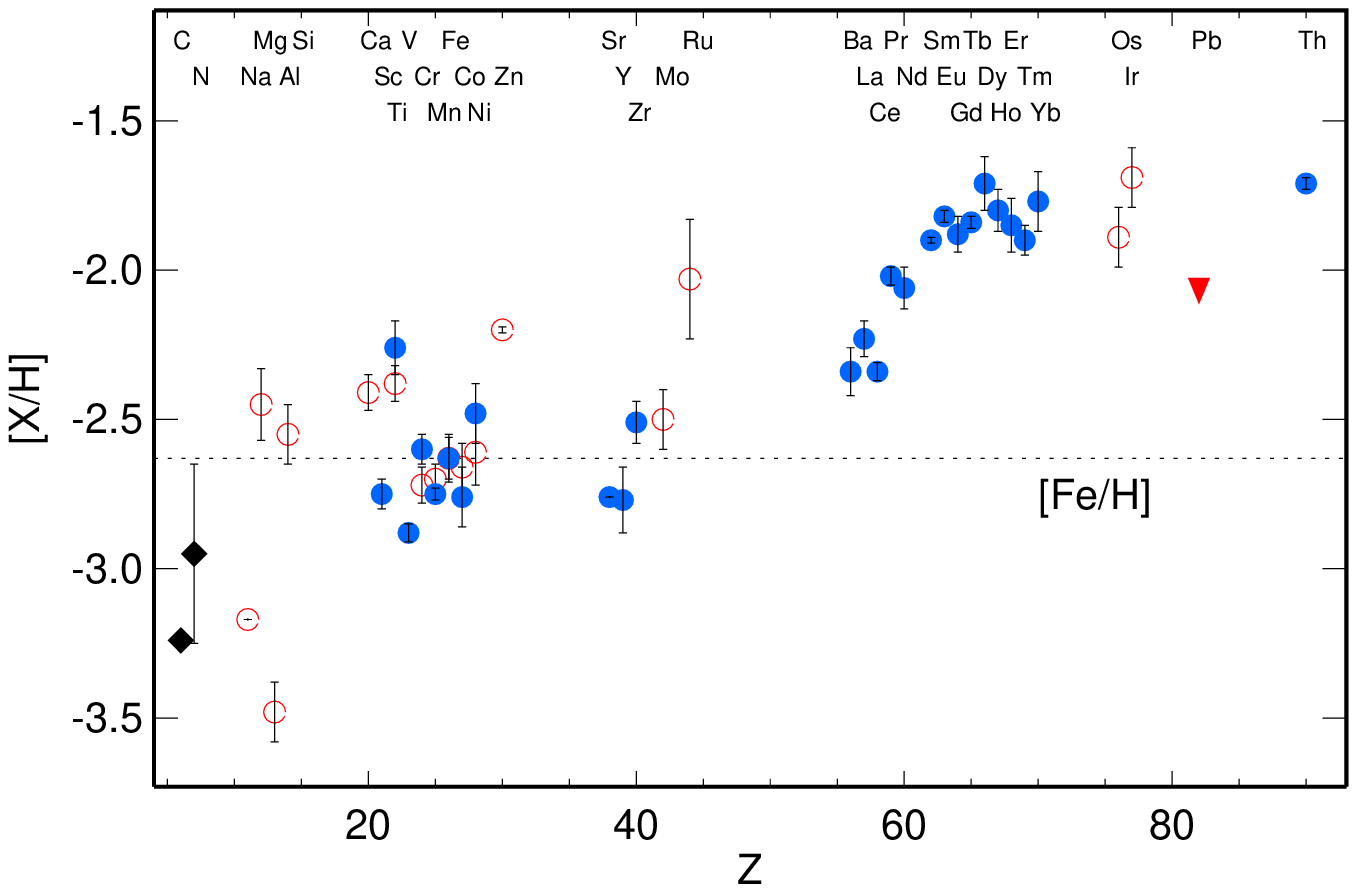}}
  \resizebox{88mm}{!}{\includegraphics{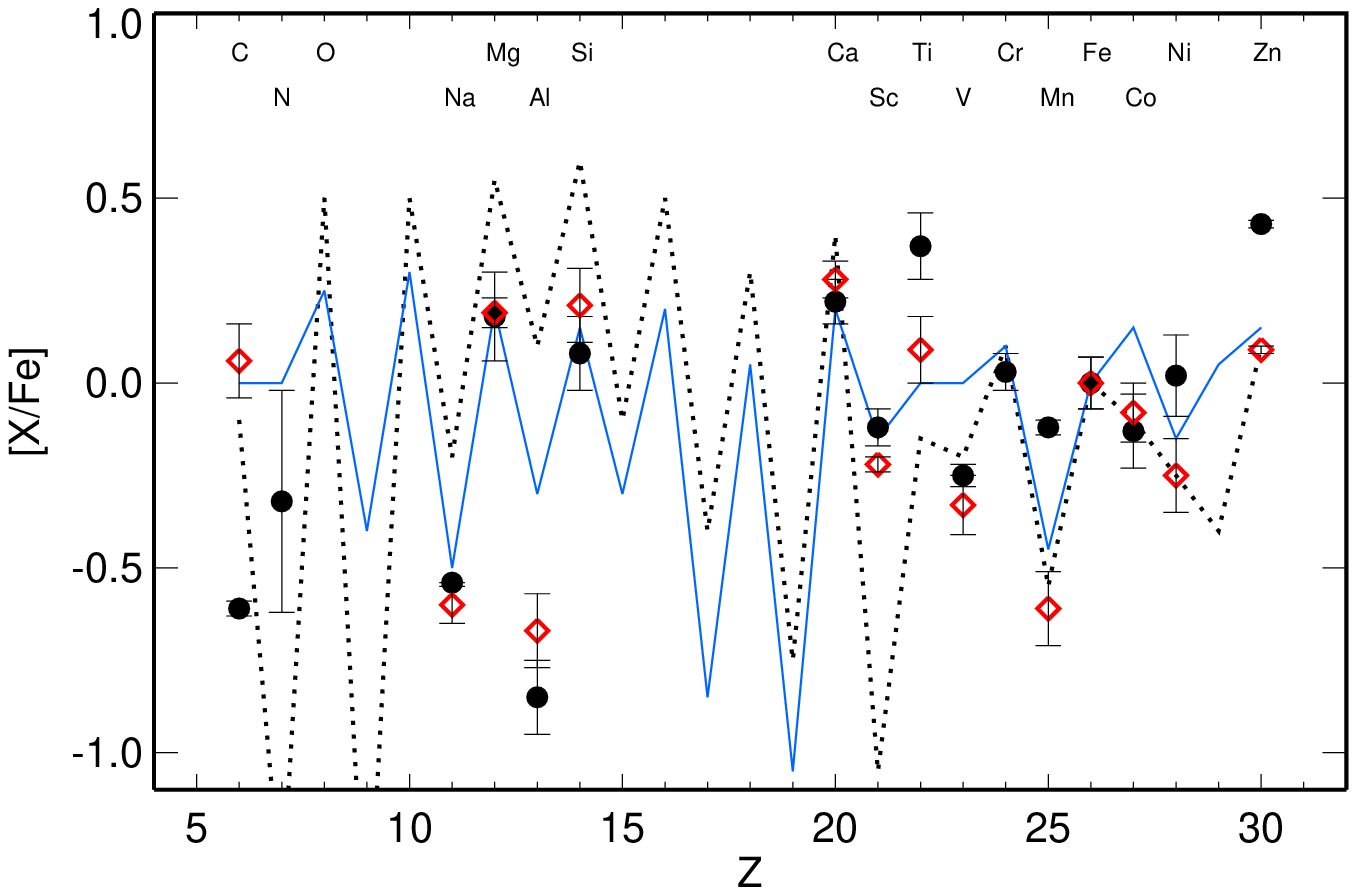}}
  \caption{\label{Fig:he2252_pattern} Left panel: the element abundance pattern of {\LyudaStar}. The filled rhombi, open circles, and filled circles display abundances derived from lines of molecules, neutral species, and singly ionised species, respectively. An upper limit for the Pb abundance is indicated by downward-facing triangle. The dotted line indicates the star's iron abundance [Fe/H] = $-2.63$. Right panel: abundance patterns of {\LyudaStar} (filled circles) and {\LyudmilasStar} (open rhombi) in the C-Zn range compared with the yields of single supernova of 14.4 $M_\odot$ (continuous line) as given by \citet{Lai2008} and 
 the chemical evolution predictions of \citet[][dotted line]{2011MNRAS.414.3231K}. The first model provides [(C+N)/Fe]. The corresponding value was plotted for {\LyudmilasStar}, but not {\LyudaStar}, where the uncertainty in derived N abundance is large.}
\end{figure*}

Owing to the high-quality and broad wavelength coverage of the spectra used in this study, we derived the abundance of 38 elements from C to Th in {\LyudaStar} and, for 22 elements of them, in the nuclear charge range between $Z = 38$ and 90. 
The oxygen, copper, and hafnium abundances could not be determined from the available observed spectrum.   
We were unsuccessful in obtaining abundances for Rh and Pd, because the strongest lines of these elements, \ion{Rh}{i} 3434\,\AA\ and \ion{Pd}{i} 3404\,\AA, could not be extracted from the noise in the $S/N \simeq 15$ observed spectrum of {\LyudaStar}.
The element abundances obtained from individual lines are listed in Table~\ref{linelist} (online material). For each feature, we provide the LTE
abundance and, for selected species, also the NLTE abundance. 
Table~\ref{Tab:AbundanceSummary} presents the mean abundances, the
number of used lines, $N_{lines}$, and $\sigma_{\eps{}}$, where $N_{lines} > 1$.
We also list the
solar photosphere abundances, $\eps{\sun}$, adopted from \citet{Lodders2009}, and the
abundances relative to iron, [X/Fe]. For computing [X/Fe],
[Fe/H]$_{\rm NLTE}= -2.63$ was chosen as the reference, with the exception of
the neutral species calculated based on a LTE assumption, where the reference
is [Fe~I/H]$_{\rm LTE}= -2.89$. Figure~\ref{Fig:he2252_pattern} displays the element abundance pattern of {\LyudaStar}.

\begin{table} 
  \renewcommand\thetable{5}
 \caption{\label{Tab:AbundanceSummary} Summary of the abundances of 
   {\LyudaStar}.}
 \centering
 \begin{tabular}{rllrrcr}\hline\hline
   $Z$ & Species & $\eps{\sun}$ & $N_{lines}$ & $\eps{ }$~~ & $\sigma_{\eps{}}$ & $\mathrm{[X/Fe]}$  \\
\noalign{\smallskip} \hline \noalign{\smallskip}
  3 & Li I  & 1.10   &  1   & $\le -0.1$  & --   & $     $\\
  6 & CH    &   8.39 &    4 &   5.15~~ &  0.02 &  --0.61 \\
  7 & NH    &   7.86 &    1 &   4.91~~ &       &  --0.32 \\
 11 & Na I  &   6.30 &    2 &   3.13$^N$ &  0.00 & --0.54 \\
 12 & Mg I  &   7.54 &    3 &   5.09$^N$ &  0.12 &  0.18 \\
 13 & Al I  &   6.47 &    1 &   2.99$^N$ &       & --0.85 \\
 14 & Si I  &   7.52 &    1 &   4.97$^N$ &       &  0.08 \\
 20 & Ca I  &   6.33 &   10 &   3.92$^N$ & 0.06  &  0.22\\  
 21 & Sc II &   3.07 &    4 &   0.32~~ &  0.05 & --0.12 \\
 22 & Ti I  &   4.90 &   10 &   2.26~~ & 0.06  &  0.25\\  
 22 & Ti II &   4.90 &   16 &   2.64~~ & 0.09  &  0.37\\
 23 & V  II &   4.00 &    3 &   1.12~~ &  0.03 & --0.25 \\
 24 & Cr I  &   5.64 &    5 &   2.66~~ & 0.06  & --0.09\\  
 24 & Cr II &   5.65 &    4 &   3.05~~ &  0.05 &  0.03 \\
 25 & Mn I  &   5.37 &    3 &   2.41~~ &  0.05 & --0.07 \\
 25 & Mn II &   5.37 &    3 &   2.62~~ &  0.02 & --0.12 \\
 26 & Fe I  &   7.45 &   43 &   4.82$^N$ & 0.08 & 0.00 \\
 26 & Fe II &   7.45 &   12 &   4.82$^N$ & 0.07 & 0.00 \\
 27 & Co I  &   4.92 &    3 &   2.00~~ &  0.08 & --0.03 \\
 27 & Co II &   4.92 &    1 &   2.16~~ &       & --0.13 \\
 28 & Ni I  &   6.23 &    6 &   3.36~~ &  0.11 &  0.02 \\
 28 & Ni II &   6.23 &    1 &   3.75~~ &       &  0.15 \\
 30 & Zn I  &   4.62 &    2 &   2.42$^T$ & 0.01 & 0.43 \\
 38 & Sr II &   2.92 &    2 &   0.16~~ &  0.00 & --0.13 \\
 38 & Sr II &   2.92 &    2 &  0.20$^N$ &  0.05 & --0.09 \\
 39 & Y  II &   2.21 &    9 & $-0.56$~~ &  0.11 & --0.14 \\
 40 & Zr II &   2.58 &   12 &    0.07~~ &  0.07 &  0.12 \\
 42 & Mo I  &   1.92 &    1 & $-0.84$~~ &       & 0.13 \\
 44 & Ru I  &   1.84 &    1 & $-0.45$~~ &       & 0.60 \\
 56 & Ba II &   2.17 &    3 & $-0.17$~~ &  0.08 &  0.29 \\
 56 & Ba II &   2.17 &    3 & $-0.32^N$ &  0.04 &  0.14 \\
 57 & La II &   1.14 &    8 & $-1.09$~~ &  0.06 &  0.40 \\
 58 & Ce II &   1.61 &   10 & $-0.73$~~ &  0.03 &  0.29 \\
 59 & Pr II &   0.76 &    3 & $-1.26$~~ &  0.03 &  0.61 \\
 60 & Nd II &   1.45 &   19 & $-0.61$~~ &  0.07 &  0.57 \\
 62 & Sm II &   1.00 &    5 & $-0.90$~~ &  0.01 &  0.73 \\
 63 & Eu II &   0.52 &    4 & $-1.30$~~ &  0.02 &  0.81 \\
 63 & Eu II &   0.52 &    4 & $-1.20^N$ &  0.03 &  0.91 \\
 64 & Gd II &   1.11 &    4 & $-0.77$~~ &  0.06 &   0.75 \\
 65 & Tb II &   0.28 &    3 & $-1.56$~~ &  0.00 &   0.79 \\
 66 & Dy II &   1.13 &   14 & $-0.58$~~ &  0.09 &   0.92 \\
 67 & Ho II &   0.51 &    4 & $-1.29$~~ &  0.07 &   0.83 \\
 68 & Er II &   0.96 &    8 & $-0.89$~~ &  0.09 &   0.78 \\
 69 & Tm II &   0.14 &    5 & $-1.76$~~ &  0.05 &   0.73 \\
 70 & Yb II &   0.86 &    1 & $-0.91$~~ &       &   0.86 \\
 76 & Os I  &   1.45 &    1 & $-0.44$~~ &       &   0.74 \\
 77 & Ir I  &   1.38 &    2 & $-0.31$~~ &  0.10 &   0.94 \\
 82 & Pb I  &   2.00 &    1 & $\le -0.78$~~ & 0.30 & $\le -0.15$ \\
 82 & Pb I  &   2.00 &    1 & $\le -0.37^N$ & 0.30 & $\le 0.26$ \\
 90 & Th II &   0.08 &    2 &  $-1.63$~~ &  0.02 &   0.92 \\
 90 & Th II &   0.08 &    2 &  $-1.55^N$ &  0.01 &   1.00 \\
\noalign{\smallskip}\hline \noalign{\smallskip}
\multicolumn{7}{l}{{\bf Notes.} $ ^{(N)}$ NLTE abundance;} \\
\multicolumn{7}{l}{$ ^{(T)}$ NLTE correction from \citet{Takeda2005zn}.} \\
\end{tabular}
\end{table}

\subsection{NLTE effects} 

NLTE abundances were derived for a number of chemical species, using the NLTE methods from \citet[][\ion{Na}{i}]{alexeeva_na}, \citet[][\ion{Mg}{i}]{mash_mg13}, \citet[][\ion{Al}{i}]{Baumueller_al1}, \citet[][\ion{Si}{i}]{Shi_si_sun}, \citet[][\ion{Ca}{i}]{mash_ca}, \citet[][\ion{Sr}{ii}]{1997ARep...41..530B}, \citet[][\ion{Ba}{ii}]{Mashonkina1999}, \citet[][\ion{Eu}{ii}, updated]{mash_eu}, and \citet[][\ion{Pb}{i}, \ion{Th}{ii}]{Mashonkina_pb}. Quantum-mechanical rate coefficients from \citet{barklem2010_na} and \citet{mg_hyd2012} were applied for Na+H and Mg+H collisions, respectively. For the remaining NLTE species, collisions with \ion{H}{i} atoms were computed with the Drawinian rates \citep{Drawin1968,Drawin1969} scaled by a factor of \kH\,= 0.1. An exception is \ion{Sr}{ii} and \ion{Ba}{ii}, for which \kH\,= 0.01 was adopted, as recommended by \citet{Mashonkina2001sr}. 

We calculated that the NLTE abundances derived from the
\ion{Na}{i}\,5890 and 5896\,{\AA} lines are $-0.44$\,dex lower than the corresponding LTE values. In contrast, the NLTE abundance correction, $\Delta_{\rm NLTE} = \eps{NLTE} - \eps{LTE}$, is positive for lines of \ion{Mg}{i}, \ion{Al}{i}, and \ion{Ca}{i}. For \ion{Ca}{i}, $\Delta_{\rm NLTE}  = 0.24$\,dex, on average.
The departures from LTE are small for \ion{Mg}{i}\,4703 and 5528\,{\AA}, with $\Delta_{\rm NLTE}$  = 0.03 and 0.01\,dex, respectively. A small negative correction of $\Delta_{\rm NLTE}  = -0.02$\,dex was calculated for \ion{Si}{i}\,3905\,{\AA}.

Evidence of notable departures from LTE was obtained for \ion{Ti}{i}, \ion{Cr}{i}, \ion{Mn}{i}, \ion{Co}{i}, and \ion{Ni}{i}. Each of the corresponding elements is observed in {\LyudaStar} for two ionisation stages, and the LTE abundance derived from lines of the neutrals is lower than that for the first ions with the following differences:

 $\Delta\eps{}$(\ion{Ti}{i} - \ion{Ti}{ii}) = $-0.38$~dex,

 $\Delta\eps{}$(\ion{Cr}{i} - \ion{Cr}{ii}) = $-0.39$~dex,

 $\Delta\eps{}$(\ion{Mn}{i} - \ion{Mn}{ii}) = $-0.21$~dex,

 $\Delta\eps{}$(\ion{Co}{i} - \ion{Co}{ii}) = $-0.16$~dex,

 $\Delta\eps{}$(\ion{Ni}{i} - \ion{Ni}{ii}) = $-0.39$~dex.

\noindent This can be understood as follows. Similar to \ion{Fe}{i}, the listed neutrals are the minority species in the atmosphere of {\LyudaStar}, and they 
are expected to be subject to ultra-violet overionisation resulting in weakened spectral lines compared with their LTE strengths, while the departures from LTE should be minor for the first ions, which are the majority species. 
For titanium, the difference in LTE abundances is largely removed, when applying the NLTE abundance corrections of 0.31~dex and 0.01~dex for \ion{Ti}{i} and \ion{Ti}{ii}, respectively, as computed by \citet{2011MNRAS.413.2184B} for the stellar parameters, $\teffm = 4600$\,K, log~g = 1.6, and [Fe/H] = $-2.5$, close to those of {\LyudaStar}. We note that the calculated NLTE effects for \ion{Ti}{i-ii} and \ion{Fe}{i-ii} are similar. In this study, we obtained $\Delta_{\rm NLTE}$ = 0.26~dex for \ion{Fe}{i} and 0.00 for \ion{Fe}{ii}. However, significantly larger NLTE corrections of 0.44\,dex and 0.64~dex were computed for \ion{Mn}{i} \citep{2008A&A...492..823B} and \ion{Co}{i} \citep{2010MNRAS.401.1334B}, respectively, using the same model, 4600/1.6/$-2.5$.
Applying these corrections 
would lead to NLTE abundances that are too high for lines of neutral Mn and Co in {\LyudaStar} compared with those for their first ions. Here, we took into consideration that $\Delta_{\rm NLTE}  = 0.07$\,dex for \ion{Co}{ii} \citep{2010MNRAS.401.1334B} and assumed that $\Delta_{\rm NLTE}  < 0.1$\,dex for \ion{Mn}{ii}. For \ion{Cr}{i-ii}, the NLTE calculations were performed for the MP dwarf stars \citep{2010A&A...522A...9B}, but not giants. In the [Fe/H] $\simeq -2.5$ stars, $\Delta_{\rm NLTE}$ is of up to 0.35~dex for \ion{Cr}{i} and 0.08~dex for \ion{Cr}{ii}. Larger positive corrections are required for the \ion{Cr}{i} lines to remove abundance discrepancy between \ion{Cr}{i} and \ion{Cr}{ii} in {\LyudaStar}. 

For \ion{Zn}{i}\,4810\,{\AA} in {\LyudaStar}, the NLTE correction is expected to be small with $\Delta_{\rm NLTE} = -0.02$~dex \citep{Takeda2005zn}. We assume that the departures from LTE for the second zinc line, \ion{Zn}{i}\,4722\,{\AA}, are as small, as is the case for \ion{Zn}{i}\,4810\,{\AA}. 
The NLTE effects for the heavy elements beyond the Fe group are discussed in Sect.\,\ref{Sect:heavy}.

In this study, we did not consider the influence of the 3D effects on the derived element abundances, although we took care to minimise such an influence, where possible, for example by removing low-excitation lines of \ion{Fe}{i} and \ion{Mn}{i}. Using the 3D hydrodynamic model atmosphere of VMP cool giant 5020/2.5/$-3$, \citet{2013A&A...559A.102D} predicted that the (3D-1D) abundance corrections are negative and large in absolute value of up to $-0.6$ to $-0.75$~dex for the lines arising from the ground state of the minority species, such as \ion{Mg}{i}, \ion{Ti}{i}, and \ion{Ni}{i}, but they decrease rapidly towards higher excitation energy of the lower level and do not exceed $-0.2$~dex, when \Eexc\ = 2~eV. We did not determine such a large discrepancy between the \Eexc\ $< 0.2$ and \Eexc\ $> 2$~eV lines for \ion{Mg}{i} and \ion{Ni}{i} (Table\,\ref{linelist}, online material). 
In contrast to the predictions, the measured difference in 1D-LTE abundances (\ion{Ti}{i} - \ion{Ti}{ii}) is negative, but not positive, and it is fully removed, when applying the 1D-NLTE abundance corrections. These results may cast a shadow of doubt on 3D calculations. However, our interpretation is that one cannot simply add 1D-NLTE and 3D-LTE results when both NLTE and 3D effects are significant.
This is exactly the case in {\LyudaStar}.

The heavy elements beyond the Fe group are mostly observed in {\LyudaStar} for their majority species, such as \ion{Ba}{ii} and \ion{Eu}{ii}, and in the low-excitation lines. \citet{2013A&A...559A.102D} predicted that the (3D-1D) corrections for the \Eexc\ = 0 lines of \ion{Ba}{ii} and \ion{Eu}{ii} are small and very similar in magnitude, $-0.05$~dex and $-0.04$~dex, respectively.

We comment below on abundances of individual groups of
elements. The sample of cool giants from \citet[][hereafter, Cayrel2004]{Cayrel2004} was chosen as our comparison sample.

\subsection{Lithium to zinc}\label{Sect:CNO}

{\it Lithium.} As expected for a red giant, {\LyudaStar} has a very low abundance of Li. The central depth of the \ion{Li}{i} 6708\,{\AA} line is about 1\,\%\ in the $S/N \simeq 100$ observed spectrum, and only an upper limit of $\eps{Li} = -0.1$ was estimated. 

{\it Carbon and nitrogen.} The C abundances obtained from CH lines in the regions 4310--4314\,{\AA} and
4362--4367\,{\AA} are consistent with each other to within 0.03\,dex (see
Table~\ref{linelist}, online material). The mean abundance is $\mathrm{[C/Fe]}
= -0.61$, which is similar to those of the giants with $\teffm < 4800$\,K
from the sample of Cayrel2004. Such a low [C/Fe] ratio in the cool giants is likely due to mixing, which has brought processed material to the surface from deep layers, where C is converted into N. This explains a factor of 2 higher N/C ratio in {\LyudaStar} compared with the solar value. The abundance of nitrogen could only be determined from the NH line at 3416.64\,\AA. We checked the spectral region around 3416.64\,{\AA} in the solar spectrum
\citep{Atlas} and fitted it with $gf$-value of the NH
line that had been reduced by $-0.4$\,dex compared with what was calculated
by \citet{Kurucz1993}. With this correction,
we derived the relative abundance [N/Fe] $= -0.32$. It is worth noting that several papers applied ``solar oscillator strengths'' of the NH molecular lines, where \citet{Kurucz1993} $\log gf$-values were reduced by a factor of 0.807~dex \citep{hill2002}, 0.4~dex \citep{Aoki_he1327,HE1219,HE2327}, and 0.3~dex \citep{Shavrina1996,Johnson2007_NH}.

{\it The $\alpha$-process elements} Mg, Si, Ca, and Ti are
enhanced relative to iron: $\mathrm{[Mg/Fe]} = 0.18$, $\mathrm{[Si/Fe]} =
0.08$, $\mathrm{[Ca/Fe]} = 0.22$, and $\mathrm{[Ti/Fe]} = 0.37$ (from \ion{Ti}{ii} lines). This is
in line with the behaviour of other metal-poor halo stars, although the [Mg/Fe] and [Ca/Fe] abundance ratios of {\LyudaStar} are significantly lower than [Mg/Fe] = 0.61 and [Ca/Fe] = 0.50 obtained by \citet{2010A&A...509A..88A} and \citet{2012A&A...541A.143S} for the Cayrel2004 stellar sample when taking departures from LTE for \ion{Mg}{i} and \ion{Ca}{i} into account. The reason, most probably, is the use of different methods for determining surface gravity and iron abundance in \citet{Cayrel2004} and this paper. This is illustrated well by the results for HD\,122563, which was included in the Cayrel2004 sample and
which was also investigated in our earlier papers \citep{Mashonkina2008,mash_fe} applying the same methods as in this study.
The absolute NLTE abundances of HD\,122563 for magnesium, $\eps{Mg}$ = 5.36, and calcium, $\eps{Ca}$ = 4.04, as determined by \citet{Mashonkina2008}, are similar to $\eps{Mg}$ = 5.39 from \citet{2010A&A...509A..88A} and  $\eps{Ca}$ = 3.97 from \citet{2012A&A...541A.143S}. However, very different iron abundances of HD\,122563 were determined in different studies. With $\teffm = 4600$\,K, \citet{Cayrel2004} derived log~g = 1.1 and [Fe/H] = $-2.82$ from LTE analysis of \ion{Fe}{i} and \ion{Fe}{ii} lines, while \citet{Mashonkina2008} calculated log~g = 1.50 from the {\sc HIPPARCOS} parallax of HD\,122563 and [Fe/H] = $-2.53$ from LTE analysis of \ion{Fe}{ii} lines.  
\citet{mash_fe} have shown that NLTE leads to consistent iron abundances for the two ionisation stages, with [Fe/H] = $-2.56$, when employing log~g = 1.60 based on the updated {\sc HIPPARCOS} parallax of HD\,122563. It is worth noting that all three VMP giants with NLTE based gravity, metallicity, and element abundances available, i.e. HD\,122563 \citep{Mashonkina2008}, {\LyudmilasStar} \citep{HE2327}, and {\LyudaStar} (this study) reveal moderate $\alpha$-enhancement with [Mg/Fe] = 0.18 to 0.31 and [Ca/Fe] = 0.21 to 0.28. 

{\it Sodium and aluminium.} {\LyudaStar} displays an underabundance of the odd$-Z$ elements Na and Al
relative to iron of $\mathrm{[Na/Fe]}=-0.54$ and
$\mathrm{[Al/Fe]}= -0.85$, and also relative to the even$-Z$ element Mg of $\mathrm{[Na/Mg]}=-0.72$ and
$\mathrm{[Al/Mg]}= -1.03$. This is not exceptional for a metal-poor halo star. From a NLTE analysis of the dwarf and giant stars in the range $-3.6 < \mathrm{[Fe/H]} < -2.5$, \citet{2010A&A...509A..88A} determined the mean values $\mathrm{[Na/Mg]} \simeq -0.8$ and $\mathrm{[Al/Mg]} \simeq -0.7$. The reason for the lower [Al/Mg] ratio of {\LyudaStar} compared with that of the VMP stellar sample is, probably, the smaller positive NLTE correction computed in this study for the \ion{Al}{i} line. For example, Fig.\,2 from \citet{2008A&A...481..481A} displays $\Delta_{\rm NLTE}$(\ion{Al}{i}\,3961\,\AA) = 0.45~dex for $\teffm = 4700$\,K, log~g = 1.65, and [Fe/H] = $-2.5$, while we compute $\Delta_{\rm NLTE}$(\ion{Al}{i}\,3961\,\AA) = 0.14~dex for the 4710/1.65/$-2.66$ model. 

\begin{figure}
\flushleft 
  \resizebox{88mm}{!}{\includegraphics{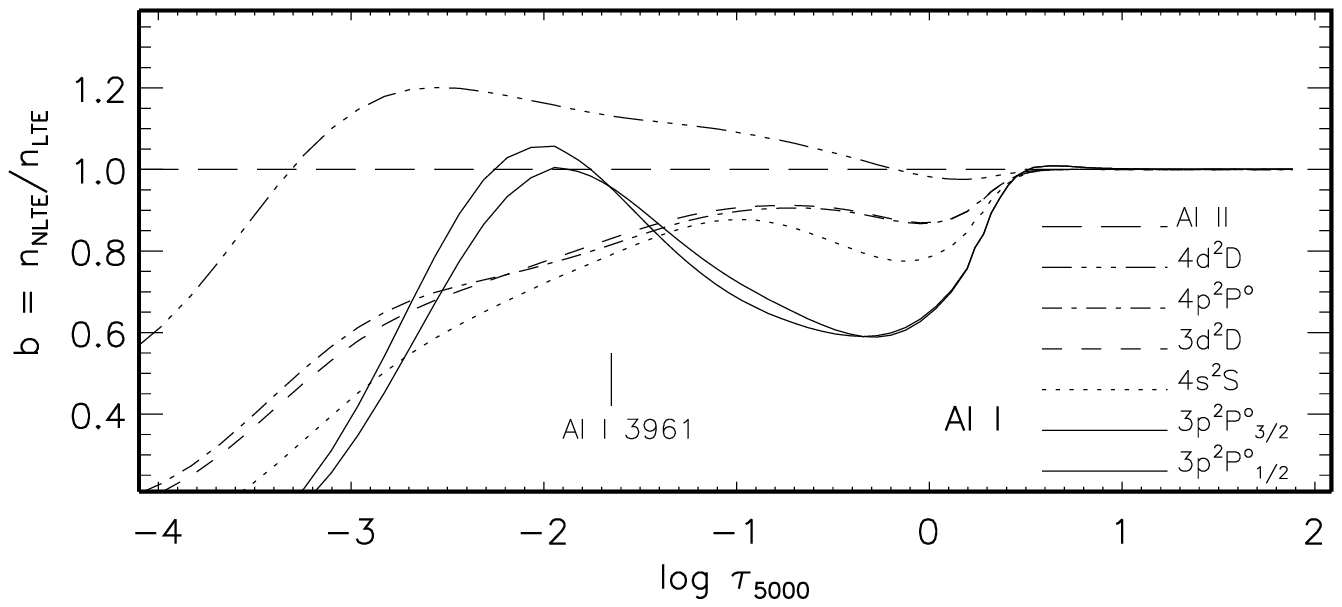}}

\flushleft 
\vspace{-7mm}
  \resizebox{88mm}{!}{\includegraphics{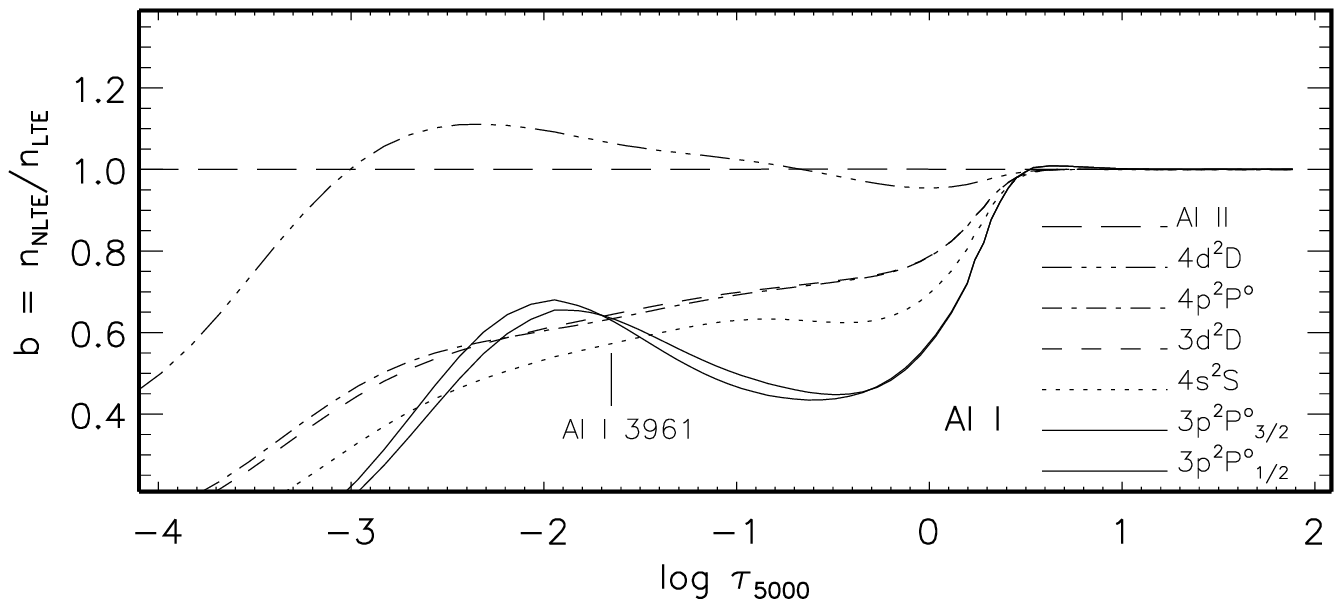}}
\caption{\label{fig:bf_al1} Departure coefficients for selected levels of
  \ion{Al}{i} as a function of
  $\log \tau_{5000}$ in the model atmosphere 4710/1.65/$-2.66$ from the calculations with the standard DETAIL opacity package (top panel) and with no lines of Fe and Ca and no quasi-molecular hydrogen absorption (bottom panel). Ticks
indicate the locations of line-centre optical depth unity for the \ion{Al}{i} line 3961\,\AA\ (transition \eu{3p}{2}{P}{\circ}{3/2} - \eu{4s}{2}{S}{}{1/2}).}
\end{figure}

We could not make the source(s) of the discrepancy clear. \citet{2008A&A...481..481A} employed a modified version of the code MULTI \citep{multi}. We would not like to cast any shadow of doubt on the modification made by \citet{1999A&A...351..168K}. Extensive tests of \citet{Bergemann_fe_nlte} for \ion{Fe}{i}-\ion{Fe}{ii} demonstrate very similar behaviour of departure coefficients computed with the codes MULTI \citep{multi} and DETAIL \citep{detail}, when using common input data. \citet{Bergemann_fe_nlte} noted: ``Perhaps, the only systematic effect is that MULTI predicts slightly stronger NLTE effects than DETAIL mainly due to the differences in the background opacity.'' We performed test calculations for \ion{Al}{i} with the code DETAIL by excluding the two sources from the background opacity, namely the lines of calcium and iron and quasi-molecular hydrogen absorption as described by \citet{1968ApJ...153..987D}, and found larger departures from LTE compared with that for the standard opacity package (Fig.\,\ref{fig:bf_al1}) and larger NLTE correction of $\Delta_{\rm NLTE}$ = 0.33~dex for \ion{Al}{i}\,3961\,\AA. 

{\it Scandium to nickel and zinc.} Abundances of the elements in the nuclear charge range between $Z = 21$ and 28 were reliably determined from lines of either two ionisation stages or the majority species of a given element. As in Paper~V, we did not use the \ion{Mn}{i} resonance lines to calculate the mean presented in
Table~\ref{Tab:AbundanceSummary}. In {\LyudaStar}, the abundance difference between \ion{Mn}{i}\,4033\,\AA\ and the three \ion{Mn}{i} subordinate lines amounts to $-0.36$\,dex. It is worth noting that for each element X observed in {\LyudaStar} for two ionisation stages, the relative LTE abundances [X\,I/Fe\,I] and [X\,II/Fe\,II] are consistent with each other to within the error bars. 
We found that {\LyudaStar} is slightly deficient in the odd$-Z$ elements Sc, V, Mn, and Co relative to iron and solar ratios with very similar [Sc/Fe], [Mn/Fe], and [Co/Fe] abundance ratios of $-0.12$, $-0.12$, and $-0.13$, respectively, and [V/Fe] = $-0.25$. The abundance of Cr and, most probably, Ni follows the iron one. From the only line of \ion{Ni}{ii} at 3769.46\,\AA, we obtained [Ni\,II/Fe\,II] = 0.15, using the predicted $gf$-value from \citet{K03}. The six lines of \ion{Ni}{i} led to a close-to-solar ratio of [Ni\,I/Fe\,I] = 0.02. The investigated star is overabundant in Zn relative to iron with [Zn/Fe] = 0.43. 

Lines of the first ions \ion{V}{ii}, \ion{Cr}{ii}, and \ion{Mn}{ii} were employed by \citet[][hereafter, Lai2008]{Lai2008} to derive the element abundances of the sample of 28 VMP stars. In the $-3.5 <$ [Fe/H] $< -2.5$ range, they obtained, on average, close-to-solar ratios for [V\,II/Fe] and [Cr\,II/Fe] and a decline in [Mn\,II/Fe] towards lower metallicity with [Mn\,II/Fe] $\simeq -0.2$ at [Fe/H] $\simeq -2.7$. Compared with the Lai2008 and Cayrel2004 stellar samples, {\LyudaStar} reveals similar element-to-iron abundance ratios for Cr, Mn, Ni, and Zn. However, {\LyudaStar} is slightly deficient in Sc, V, and Co relative to Fe, in contrast to stars of close metallicities in both VMP stellar samples.
 
We found that the element abundance pattern of {\LyudaStar} in the Na--Zn range (Fig.~\ref{Fig:he2252_pattern}) resembles that of the VMP giant {\LyudmilasStar} (5050/2.34/$-2.78$), which was investigated in Paper~V using the same methods as in this study. The star {\LyudaStar} has a significantly lower C abundance, presumably because it is a more evolved star compared with {\LyudmilasStar}. The difference concerns with Ti, Mn, and Zn, which have higher abundances relative to Fe in {\LyudaStar} compared with {\LyudmilasStar}. 

On the basis of their Na to Zn abundances, {\LyudaStar} and {\LyudmilasStar} do not appear to be exceptional. \citet{Lai2008} fitted the average abundance pattern of their stellar sample to the entire library of single supernova (SN) yields, attempting to define a "typical'' Population~III star that has enriched the interstellar medium out of which the VMP stars formed. A fairly good fit was obtained for SN of 14.4 $M_\odot$. In Fig.~\ref{Fig:he2252_pattern}, we compare the abundance patterns of {\LyudaStar} and {\LyudmilasStar} with the best fit model from \citet{Lai2008} and also the chemical evolution calculations of \citet{2011MNRAS.414.3231K}. 
In general, the single SN model fits the observations well. The exceptions are Al, V, and Co, which are overproduced in the model, and Ti, which, in contrast, is underproduced. The observed underabundances of V and Co relative to Fe are reproduced well by the model of \citet{2011MNRAS.414.3231K}.

\subsection{Heavy elements: Sr to Th}\label{Sect:heavy}

We measured five light trans-iron elements with $38 \le Z \le 44$, 14 elements in the region of the second $r$-process peak, osmium and iridium, which represent the third peak, and the actinide thorium. For the Pb abundance, we could only estimate an upper limit.

\begin{figure*}  
  \resizebox{\textwidth}{!}{\includegraphics{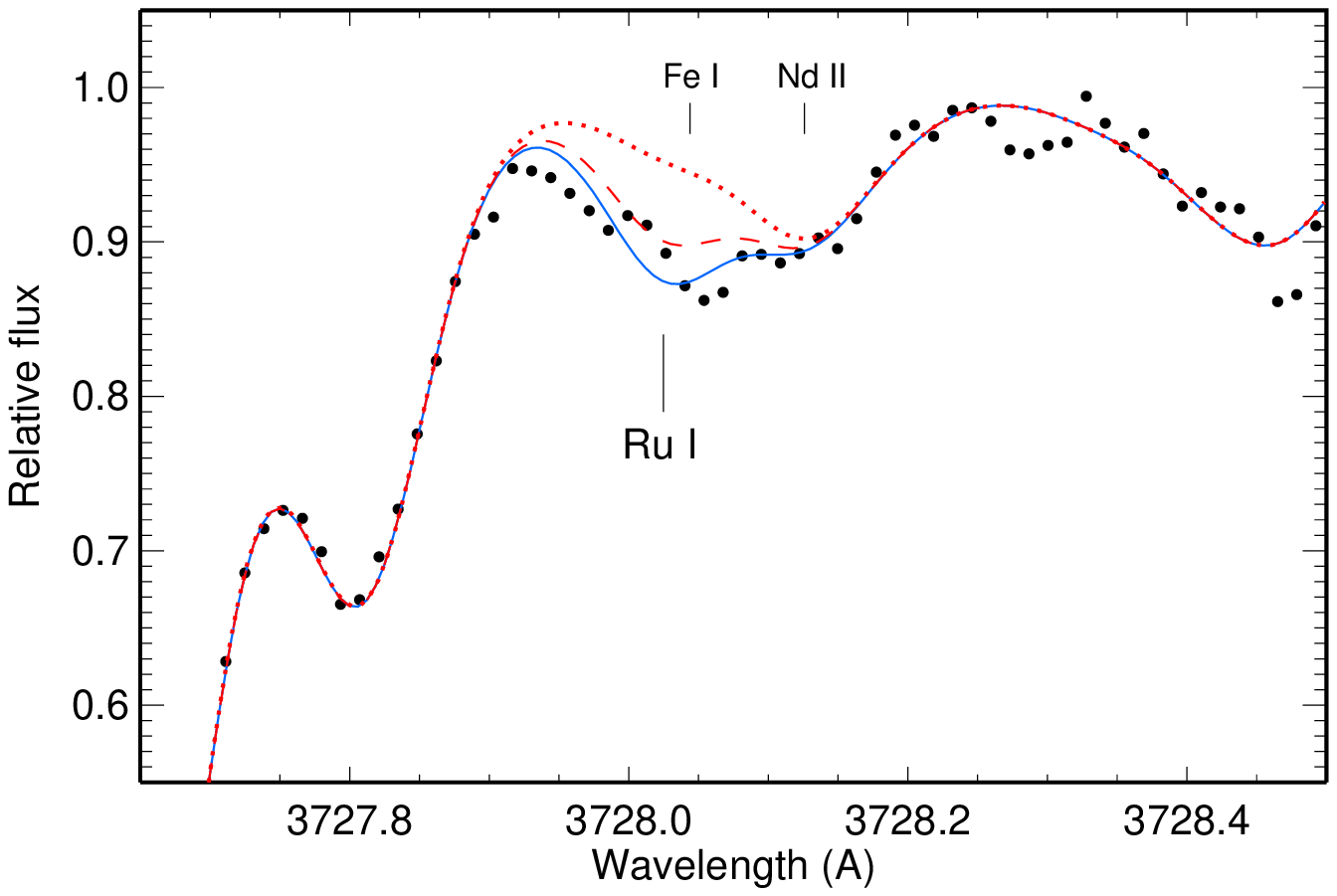}
                            \includegraphics{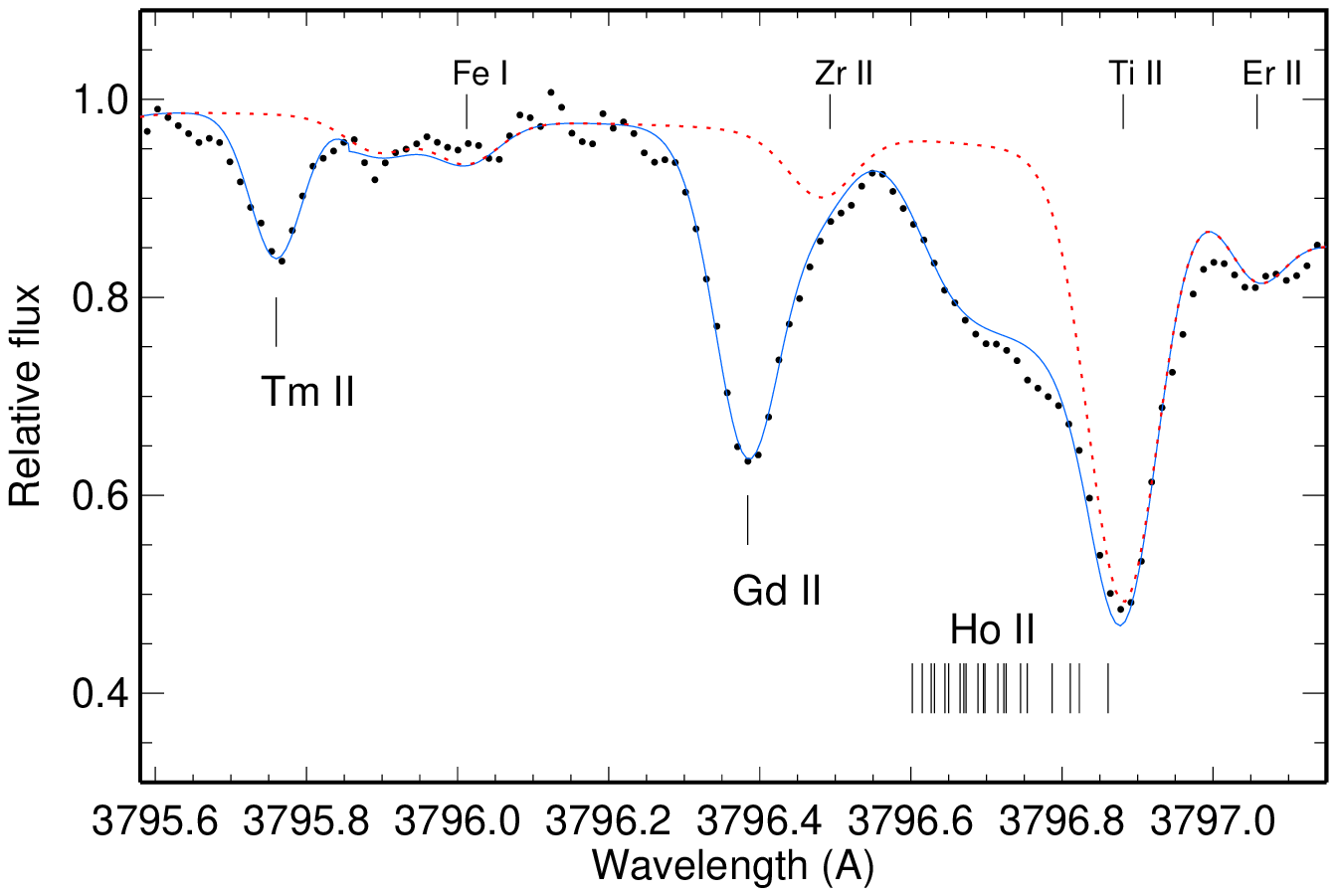}}
  \resizebox{\textwidth}{!}{\includegraphics{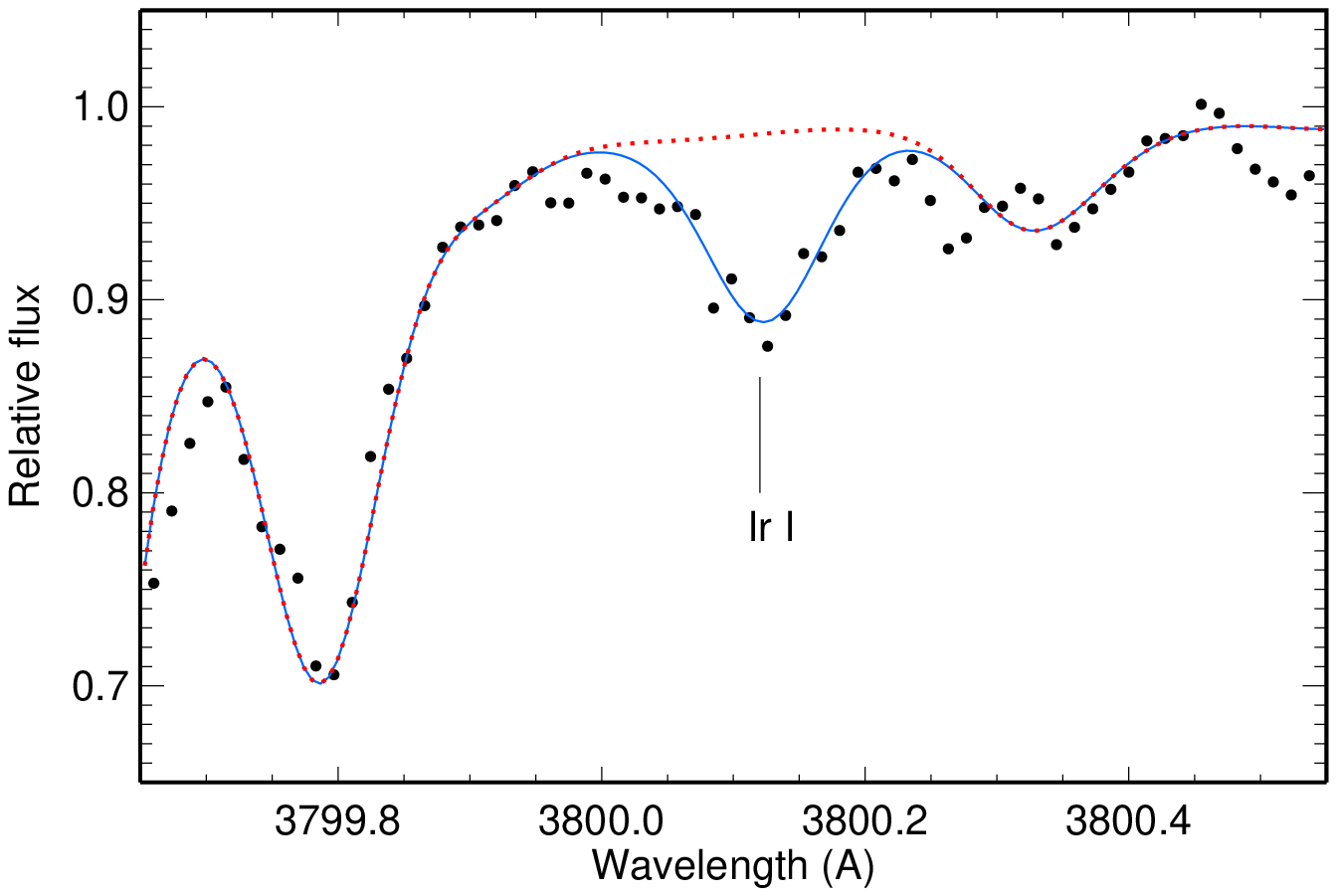}
                            \includegraphics{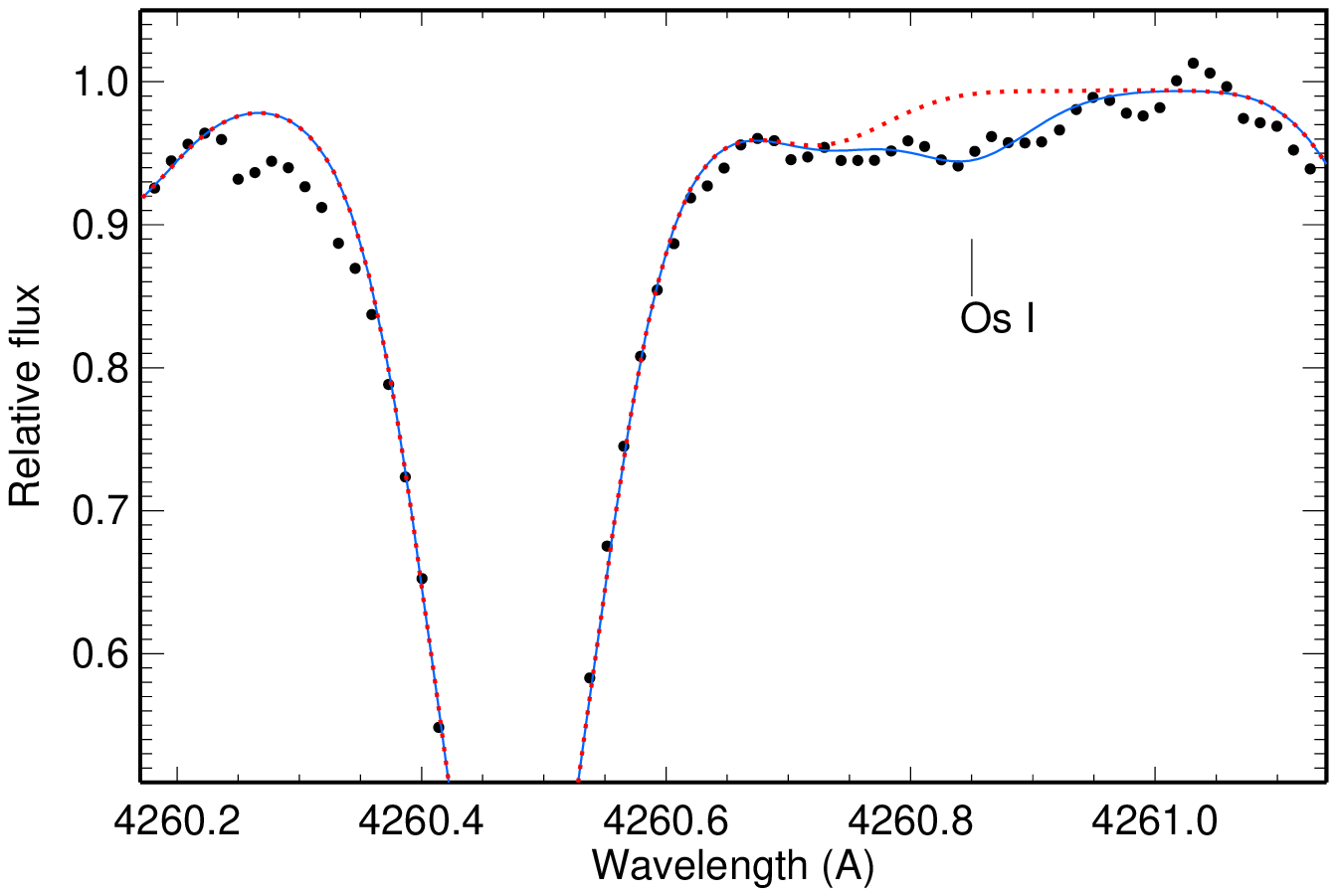}}
  \caption{\label{Fig:tm_gd_ho} Best fits (continuous curve) of the 
    spectral ranges 3727.7-3728.5\,{\AA} and 3795.6-3797.1\,{\AA} (top row), 
    where the investigated \ion{Ru}{i}, \ion{Tm}{ii}, \ion{Gd}{ii}, and 
    \ion{Ho}{ii} lines are located; \ion{Ir}{i} 3800\,{\AA} 
    and \ion{Os}{i} 4260\,{\AA} (bottom row) in the 
    observed spectrum of {\LyudaStar} (bold dots).
    The dotted curves show the synthetic
    spectrum with no relevant element in the atmosphere. The dashed
    curve in the top left panel shows the effect of a 0.2\,dex lower Ru abundance 
    on the synthetic spectrum.}
\end{figure*}

\citet{HERESIX} showed an advantage of applying the NLTE approach when constraining a pure $r$-process Ba/Eu ratio from observations of the strongly $r$-process enhanced VMP stars. In this study, NLTE abundances were determined for five heavy elements: Sr, Ba, Eu, Pb, and Th. They are presented in Tables~\ref{Tab:AbundanceSummary} and \ref{linelist} (online material). The largest departures from LTE were found for \ion{Pb}{i} 4057\,\AA. This can be understood, because \ion{Pb}{i} represents a minor fraction of the element abundance, and its number density can easily deviate from the population in thermodynamic equilibrium, because of deviations in the mean intensity of ionising radiation from the Planck function. NLTE leads to overionisation of \ion{Pb}{i} resulting in the weakened line and positive $\Delta_{\rm NLTE}$ of 0.41~dex. Different NLTE mechanisms connected with the radiative bound-bound transitions are in action for \ion{Sr}{ii}, \ion{Ba}{ii}, \ion{Eu}{ii}, and \ion{Th}{ii} that represent a major fraction of their element abundances.
The \ion{Sr}{ii} resonance lines are strong, and they are weakly affected by departures from LTE. The barium abundance of {\LyudaStar} was determined
from the three subordinate lines, \ion{Ba}{ii} 5853, 6141, and 6497\,{\AA},
which are almost free of HFS effects. Our NLTE calculations for {\LyudaStar} showed that they are stronger than in the LTE case,
resulting in a negative NLTE abundance correction of different magnitude for different lines, $\Delta_{\rm NLTE} = -0.02$\,dex, $-0.17$\,dex, and $-0.23$\,dex, respectively. 
NLTE largely removes the difference in LTE abundances between different \ion{Ba}{ii} lines, and the statistical error reduces to $\sigma_{\eps{}}$ = 0.04~dex. We calculated positive NLTE abundance corrections for the lines of 
\ion{Eu}{ii} and \ion{Th}{ii}, finding values close to $+0.1$\,dex. All the elements from lanthanum to ytterbium are observed in the lines of their majority
species, with term structures as complicated as that for \ion{Eu}{ii} and \ion{Th}{ii}, so
the departures from LTE are expected to be similar to those for \ion{Eu}{ii} and \ion{Th}{ii}. This is largely true also for osmium and iridium detected in the lines of
their neutrals, \ion{Os}{i} and \ion{Ir}{i}, which have relatively high
ionisation energies of 8.44 and 8.97~eV, respectively. Fortunately, the
abundance \emph{ratios} among heavy elements except Ba are probably only weakly affected
by departures from LTE. Indeed, the differences in mean abundance between Th and Eu are very similar in LTE and NLTE, $-0.33$\,dex and $-0.35$\,dex, respectively. 
For consistency, we used the abundances
of the heavy elements beyond strontium as determined based on the LTE assumption.

\begin{figure}  
  \resizebox{88mm}{!}{\includegraphics{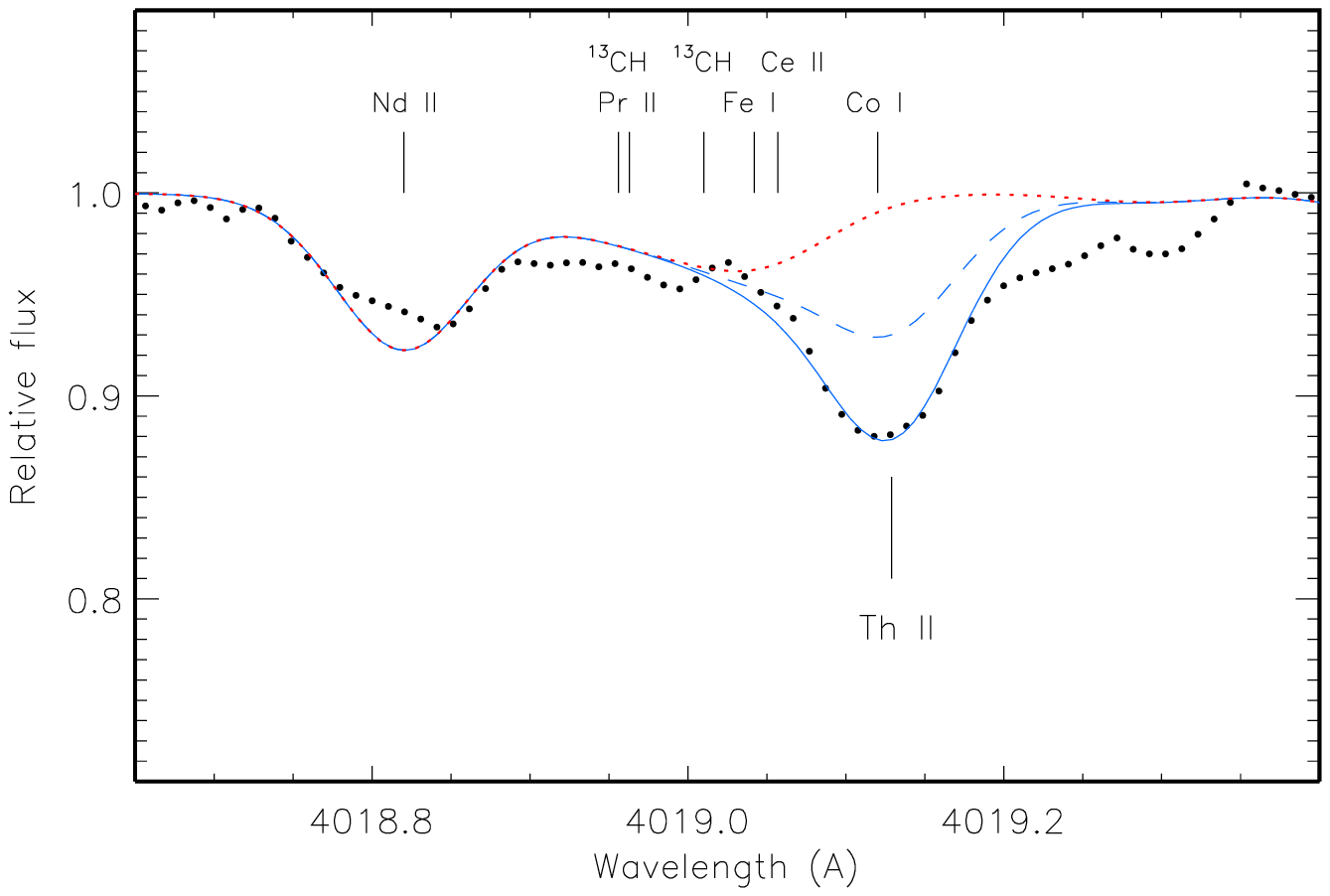}}
  \resizebox{88mm}{!}{\includegraphics{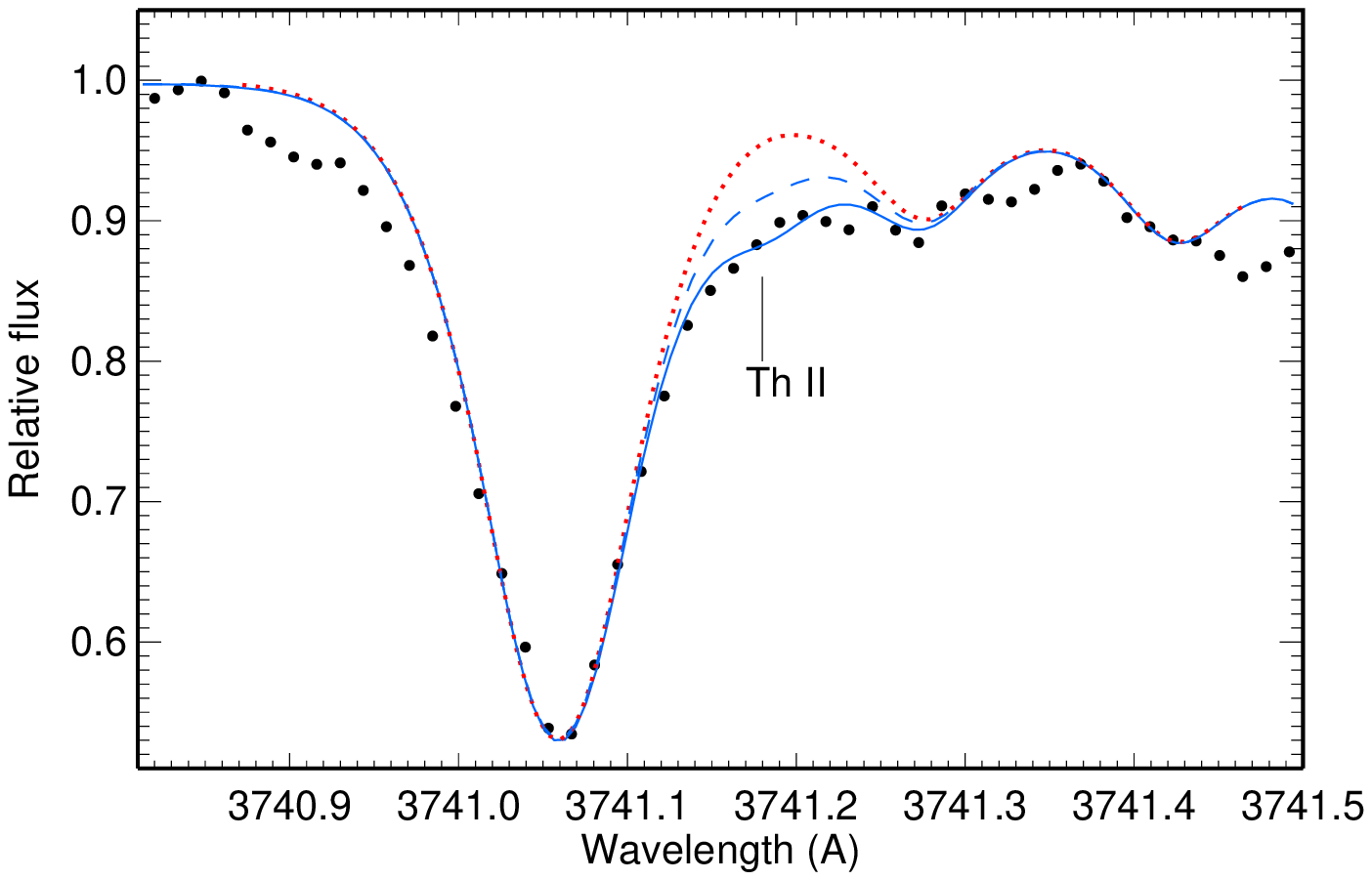}}
  \caption{\label{Fig:th_line} Same as in Fig.~\ref{Fig:tm_gd_ho} for 
    \ion{Th}{ii} 4019\,{\AA} (top panel) and \ion{Th}{ii} 3741\,{\AA} (bottom panel). The dashed
    curves show the effect of a 0.3\,dex lower Th abundance on the
    synthetic spectrum. }
\end{figure}

The abundance analysis was based on line profile fitting. Figures~\ref{Fig:tm_gd_ho} and \ref{Fig:th_line} illustrate the quality of the fits for a few selected lines. For most chemical species, 2 to 19 spectral lines were employed, and nowhere does $\sigma_{\eps{}}$ exceed 0.11~dex. For Mo, Ru, Yb, and Os, the abundance was derived from a single  line. \ion{Mo}{i} 3864\,\AA, \ion{Yb}{ii} 3694\,\AA, and \ion{Os}{i} 4260\,\AA\ (Fig.~\ref{Fig:tm_gd_ho}, bottom right panel) were found to be free of blends in the spectrum of {\LyudaStar}. Because of the high 
$S/N\simeq 100$ of the observed spectrum in the relevant wavelength range, the uncertainty in the derived abundances was estimated as 0.1\,dex.
The observed feature at 3728.0-3728.2\,\AA\ (Fig.~\ref{Fig:tm_gd_ho}, top left panel) is attributed to a combination of the \ion{Ru}{i} 3728.025\,\AA, \ion{Fe}{i} 3728.044\,\AA, \ion{Nd}{ii} 3728.126\,\AA, and two lines of \ion{Ce}{ii} at 3728.019 and 3728.180\,\AA. For the blending lines, atomic data were taken from the VALD database \citep[][hereafter, VALD]{vald}. Ignoring the \ion{Ce}{ii} lines 
completely leads to a $0.1$\,dex higher ruthenium abundance. The uncertainty in Ru abundance grows, when considering a possible variation in the Nd and Fe abundances.  
We carefully estimate $\sigma_{\eps{}} = 0.2$\,dex for the Ru abundance obtained.

The radioactive element thorium was clearly detected in {\LyudaStar} in the two lines, \ion{Th}{ii} 4019 and 3741\,\AA\ (Fig.~\ref{Fig:th_line}). 
To determine the Th abundance as accurately as possible, we updated the list of lines contributing to the 4019\,\AA\ blend. The used lines, together with the sources of atomic data, are indicated in Table\,\ref{Tab:th_list}. 
The second line, \ion{Th}{ii} 3741.183\,\AA, is located between the strong \ion{Ti}{i} 3741.059\,\AA\ (\Eexc/log~$gf$ = 0.02~eV/$-0.21$) and \ion{Sm}{ii} 3741.276\,\AA\ (0.19~eV/$-0.59$) lines. Atomic data are indicated according to VALD. To fit the observed feature at 3741\,\AA\ in {\LyudaStar}, we used a 0.6~dex lower $gf$-value of \ion{Ti}{i} 3741\,\AA\ and
$\eps{Sm} = -0.88$, which is only 0.02~dex higher than the obtained mean Sm abundance. The abundance derived from \ion{Th}{ii} 3741\,\AA\ agrees to within 0.03~dex with that for \ion{Th}{ii} 4019\,\AA.

\begin{table} 
  \renewcommand\thetable{6}
 \caption{\label{Tab:th_list} Line list adopted for the spectrum synthesis of the 4019\,\AA\ blend.}
 \centering
 \begin{tabular}{rcccc}\hline\hline \noalign{\smallskip} 
 Species & $\lambda$(\AA) & $E_{\rm exc}$(eV) & log $gf$ & Refs. \\
\noalign{\smallskip} \hline \noalign{\smallskip} 
\ion{Ce}{ii} & 4018.820 & 1.546 & --0.960 & 1 \\
\ion{Nd}{ii} & 4018.820 & 0.064 & --0.880 & 1 \\
\ion{Fe}{i}  & 4018.887 & 4.256 & --2.604 & 1 \\ 
\ion{Ce}{ii} & 4018.900 & 1.013 & --1.220 & 1 \\ 
\ion{Ce}{ii} & 4018.927	& 0.635 & --1.680 & 1 \\
\ion{V}{i}   & 4018.929 & 2.581 & --0.557 & 1 \\
 $^{13}$CH   & 4018.956 & 0.463 & --1.379  & 2 \\
\ion{Pr}{ii} & 4018.963 & 0.204 & --1.030 & 1 \\ 
\ion{Mn}{i}  & 4018.987 & 4.354 & --1.884 & 1 \\  
\ion{Fe}{i}  & 4019.003 & 4.320 & --1.793 & 1 \\   
 $^{13}$CH   &  4019.010 & 0.463 & --1.354 & 2 \\    
\ion{Fe}{i}  & 4019.042 & 2.608 & --2.780 & 1 \\
\ion{V}{ii}  & 4019.044 & 3.753 & --1.232 & 1 \\     
\ion{Ce}{ii} & 4019.057 & 1.014	& --0.530 & 1 \\
\ion{Mn}{i}  & 4019.066 & 4.666 & --0.523 & 1 \\   
\ion{Ni}{i}  & 4019.067 & 1.935 & --3.121 & 1 \\ 
\ion{Co}{i}  & 4019.110 & 2.280 & --3.287 & 3 \\
\ion{Co}{i}  & 4019.118 & 2.280 & --3.173 & 3 \\   
\ion{Co}{i}  & 4019.120 & 2.280 & --3.876 & 3 \\  
\ion{Co}{i}  & 4019.125 & 2.280 & --3.298 & 3 \\ 
\ion{Co}{i}  & 4019.125 & 2.280 & --3.492 & 3 \\  
\ion{Th}{ii} & 4019.129 & 0.000 & --0.228 & 1 \\    
\ion{Co}{i}  & 4019.134 & 2.280 & --3.287 & 3 \\   
\ion{V}{i}   & 4019.134 & 1.804 & --2.234 & 1 \\     
\ion{Co}{i}  & 4019.135 & 2.280 & --3.474 & 3 \\   
\ion{Co}{i}  & 4019.138 & 2.280 & --3.173 & 3 \\     
\ion{Co}{i}  & 4019.140 & 2.280 & --3.298 & 3 \\  
\ion{Ce}{ii} & 4019.271	& 0.328 & --2.320 & 1 \\
\ion{Co}{i}  & 4019.292	& 0.580 & --4.371 & 1 \\
\ion{Co}{i}  & 4019.297	& 0.630 & --3.432 & 1 \\
\noalign{\smallskip}\hline \noalign{\smallskip}
\multicolumn{5}{l}{{\bf Refs.} 1 = VALD, 2 = \citet{hill2002},} \\
\multicolumn{5}{l}{ \ 3 = \citet{Johnson2001}.} \\
\end{tabular}
\end{table}

\begin{figure}  
  \resizebox{88mm}{!}{\includegraphics{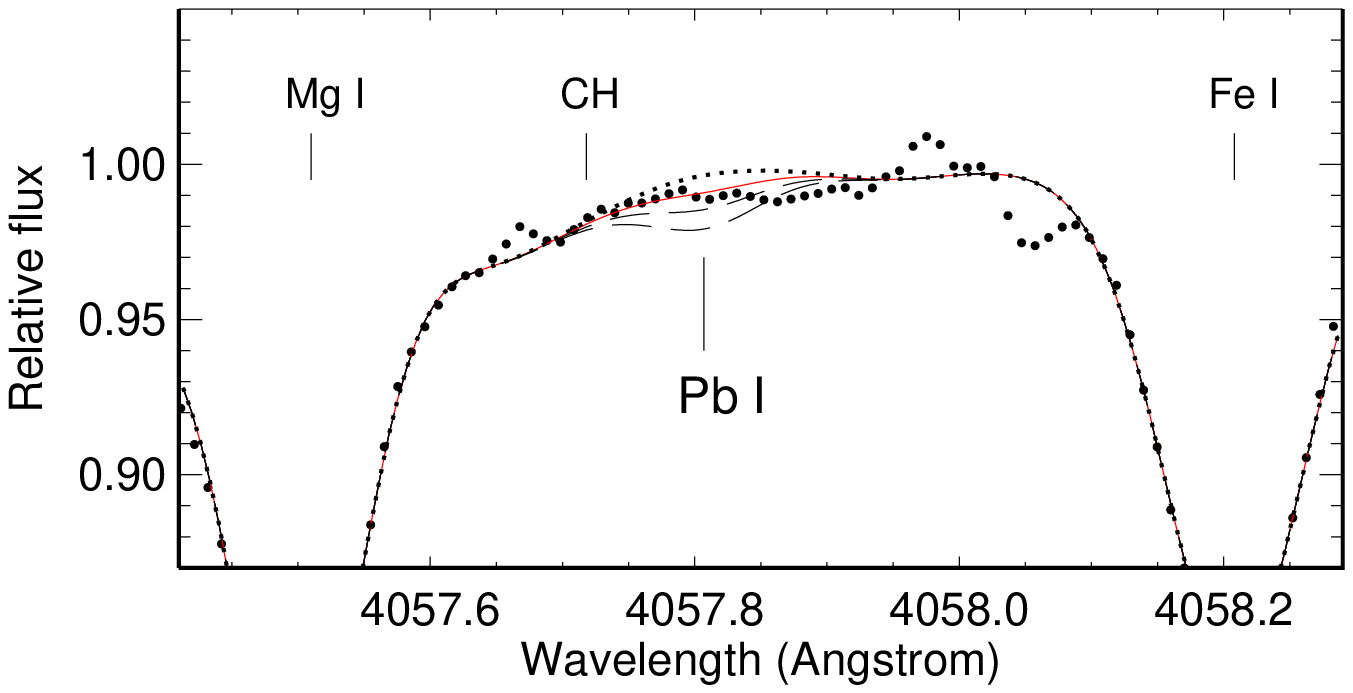}}
  \caption{\label{Fig:pb_line} $S/N \simeq 80$ spectrum of {\LyudaStar} (bold dots) in the region, where the \ion{Pb}{i} 4057\,{\AA} line is located. The synthetic spectra were computed with no Pb in the atmosphere (dotted curve), $\eps{Pb} = -0.78$ (continuous curve), $-0.48$, and $-0.28$ (dashed curves for both values).}
\end{figure}

We attempted to detect the second actinide uranium. The only useful line, \ion{U}{ii} 3859.57\,{\AA}, lies between the two strong lines, \ion{Fe}{i} 3859.21\,{\AA} and \ion{Fe}{i} 3859.91\,{\AA}, in their overlapping wings. Our calculations with [U/Fe] = 1 predicted a depression of about 1\,\%\ at the \ion{U}{ii} 3859\,{\AA} line centre, however, no absorption feature was found at this place in the observed spectrum of {\LyudaStar}. This indicates [U/Fe] $< 1$ for {\LyudaStar}, although [U/Fe] $\simeq 1$ cannot be ruled out, because the $S/N$ is about 100 in this region.

It would be of great interest to know the Pb abundance of {\LyudaStar}, since Th and U decay to the stable element lead. In the visual spectra, lead can only be observed in the two lines of similar strengh, i.e. \ion{Pb}{i} 3683\,{\AA} (\Eexc\ = 0.97\,eV, log~$gf = -0.52$) and 4057\,{\AA} (\Eexc\ = 1.32\,eV, log~$gf = -0.17$). However, the first line lies in a rather crowded spectral region, and most stellar Pb abundance analyses rely on using \ion{Pb}{i} 4057\,{\AA}. In the spectrum of {\LyudaStar}, there is a hint of a weak absorption feature at the \ion{Pb}{i} 4057\,{\AA} line centre (Fig.\,\ref{Fig:pb_line}). However, the $S/N$ is quite low ($\simeq 80$), and we could only estimate an upper limit for the Pb abundance to be $\eps{LTE} = -0.78$ and $\eps{NLTE} = -0.37$.

\section{Heavy-element abundance pattern of {\LyudaStar} and r-II stars}\label{Sect:ssr}

This study confirms that {\LyudaStar} is enhanced in the heavy elements beyond Ba relative to iron and solar ratios (Fig.~\ref{Fig:he2252_pattern}). For seven
elements (i.e. Eu, Gd, Tb, Dy, Ho, Er, and Tm), where there is an $r$-process contribution to
the SS matter of more than 80\,\% according to the predictions of
 \citet[][hereafter, Bisterzo2014]{2014arXiv1403.1764B}, the average abundance ratio is $[r/\mathrm{Fe}] = 0.80\pm0.06$. The obtained [Ba/Eu] $= -0.52$ is consistent with the earlier estimate $\mathrm{[Ba/Eu]}=-0.54$ from Paper\,II, and the difference between [Eu/Fe] = 0.81 (this study) and [Eu/Fe] = 0.99 (Paper\,II) is mostly due to the higher revised iron abundance. 

\begin{figure*} 
  \centering
  \resizebox{150mm}{!}{\includegraphics{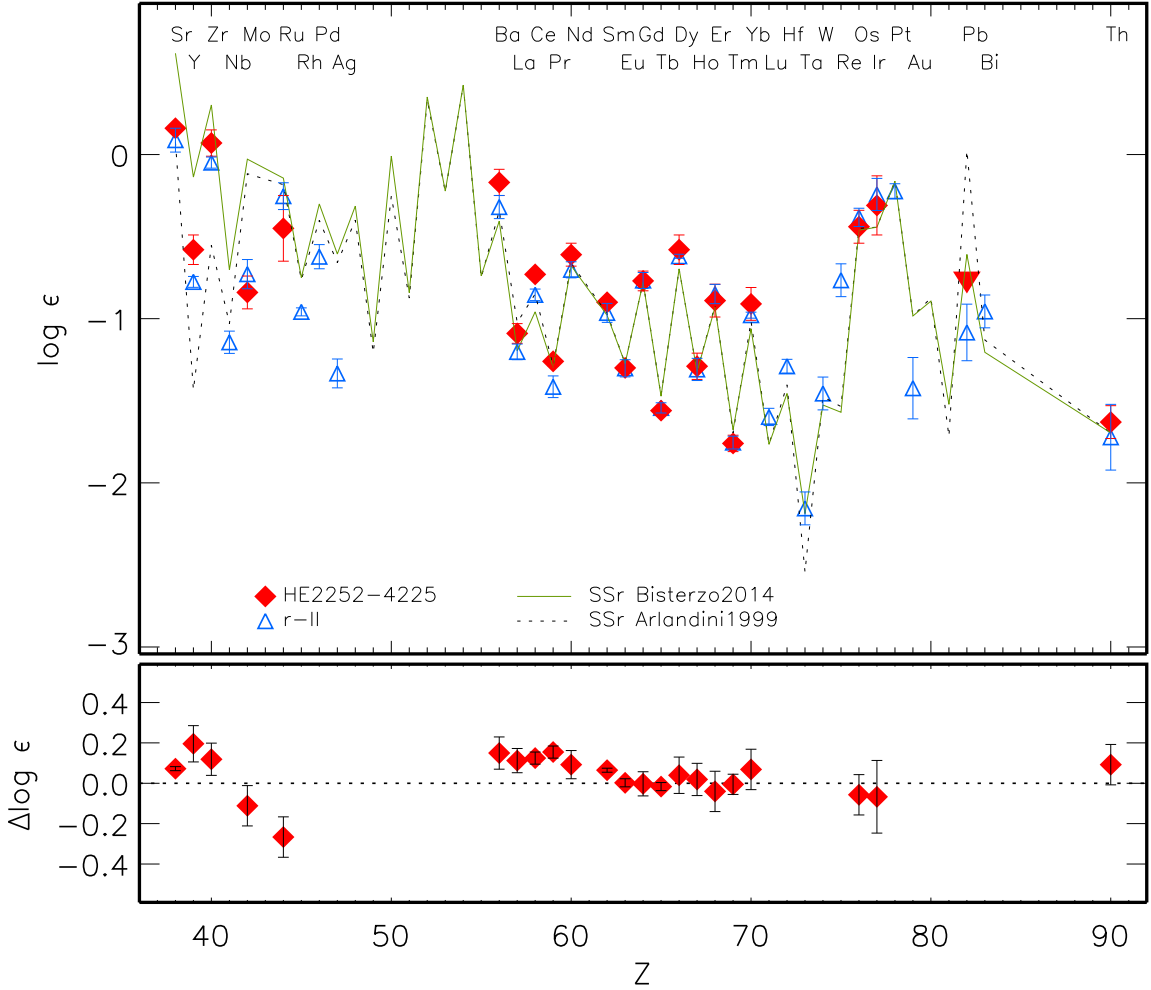}}
  \caption{\label{Fig:AbundancePattern} Heavy-element abundance pattern of 
    {\LyudaStar} (filled rhombi) compared with the average element abundances (open triangles) of the four benchmark r-II stars,
    CS\,22892-052, CS\,31082-001, HE\,1523-0901, and HE\,1219-0312. An upper limit for the Pb abundance of {\LyudaStar} is indicated by downward-facing triangle.
For the r-II stars, the error bars show the dispersion in the single star measurements around the mean, when the species was measured in more than one star. Otherwise it was adopted as 0.1~dex. For comparison, the SSr abundance patterns calculated using the $s$-process predictions of 
    \citet{Arlandini1999} (dotted curve) and \citet{2014arXiv1403.1764B} (continuous curve)
    are shown. They and also the heavy-element abundances of the r-II stars have 
    been scaled to match Eu--Tm in {\LyudaStar}. The bottom panel displays the
    difference between {\LyudaStar} and the r-II stars defined as $\Delta\eps{} =
    \eps{HE\,2252-4225} - \eps{r-II}$. }
\end{figure*}

 For the abundance comparison, we choose the four r-II stars with the largest number of measured species in the Sr--Th region. These are CS\,22892-052 \citep{Sneden2003}, CS\,31082-001 \citep{2013A&A...550A.122S}, HE\,1523-091 \citep{Sneden2008}, and HE\,1219-0312 \citep{HE1219}. For each of them, the heavy-element abundances have 
    been scaled to match Eu--Tm in {\LyudaStar}, and the scaled abundances have 
    been averaged. 
The average abundances are plotted in Fig.~\ref{Fig:AbundancePattern}. 
The dispersion in the single star measurements around the mean does not exceed 0.1~dex for every investigated species except Th, indicating very similar abundance patterns of the r-II stars in the Sr--Ir region and a common origin of the first, second, and third $r$-process peak elements in the classical $r$-process. 
It is clear from Fig.~\ref{Fig:AbundancePattern} that the elements in the range from Ba to Yb and also Os and Ir in {\LyudaStar} match the scaled abundances of the r-II stars very well, with a mean difference of $\Delta\eps{} = \eps{HE\,2252-4225} - \eps{r-II} = 0.04\pm0.07$. 
 No notable discrepancies between {\LyudaStar} and the r-II stars are also found for the light trans-iron elements Sr, Y, Zr, and Mo. For example [Sr/Eu] $= -0.94$ and [Zr/Eu] $= -0.69$ of {\LyudaStar} are in excellent agreement with the average abundance ratios [Sr/Eu] $= -0.92\pm0.13$ and [Zr/Eu] $= -0.64\pm0.19$ calculated in Paper~V for nine r-II stars, and they are significantly lower than the corresponding values, $-0.56$ and $-0.31$, for the 32 r-I stars. Thus, {\LyudaStar} should be referred to as r-II star, despite the slightly low derived europium enhancement of [Eu/Fe] = 0.81 (LTE) and 0.91 (NLTE).

As proposed by \citet{HE2327}, a membership of a given $r$-process enhanced MP star to the r-II or r-I group could be connected to an origin of the neutron-capture elements, if a separation involved the criteria based on not only [Eu/Fe] and [Ba/Eu], but also [Sr/Eu]. For example, the 
well-studied stars HD\,221170 \citep{Ivans2006} and BD\,+17$^{\circ}$3248 \citep{Cowan2002} have very similar [Eu/Fe] $= 0.80$ and 0.91 and [Ba/Eu] $= -0.54$ and $-0.51$ to those for {\LyudaStar}, and by definition (Paper~I), all three objects are of type r-I. However, [Sr/Eu] ratios of $-0.67$ and $-0.62$ for HD\,221170 and BD\,+17$^\circ$3248, respectively, are higher than for {\LyudaStar} ($-0.94$), and an
 origin of the first $r$-process peak elements in HD\,221170 and BD\,+17$^{\circ}$3248 remains unclear.

In Fig.~\ref{Fig:AbundancePattern}, we also plot the scaled solar
$r$-process residuals calculated by subtracting theoretical
$s$-process yields from the observed SS total abundance. For the latter, we rely on the meteoritic abundances of \citet{Lodders2009}. For the solar $s$-abundances, we consider the $s$-process predictions of 
    \citet[][stellar model; hereafter, Arlandini1999]{Arlandini1999} and \citet{2014arXiv1403.1764B}. Arlandini1999 used stellar AGB models of 1.5 and 3~$M_\odot$ with half solar metallicity. Bisterzo2014 performed the Galactic chemical evolution (GCE) calculations that considered the $s$-process contributions from different generations of AGB stars of various mass.
Hereafter, the solar $r$-residuals
are referred to as the solar system $r$-process (SSr) abundance pattern. 
A notable difference between two sets of the solar $r$-residuals was obtained for all elements with dominant $s$-process
contribution to their solar abundances. For example for Sr, Y, and Zr, Arlandini1999 predicted a main $s$-process contribution of 85, 92, and 83\,\%, respectively, while the corresponding numbers in Bisterzo2014 are 69, 72, and 66\%.

The elements in the range from La to Yb and also the third $r$-process peak elements Os and Ir in {\LyudaStar} were found to match
the solar $r$-process pattern very well, with a mean difference of $\Delta\eps{} = \eps{HE\,2252-4225} - {\rm SSr} = 0.03\pm0.07$ and $0.05\pm0.08$ for the solar $r$-abundances of Arlandini1999 and Bisterzo2014, respectively. 
This finding is in line with the earlier results obtained for other $r$-process rich stars \citep[for a review, see][]{Sneden2008} and provides
 additional evidence of universal production ratio of these elements during
the Galactic history. As for Ba, it does matter whether its stellar abundance and also abundances of the $r$-process elements are determined in LTE or NLTE. The difference between the barium LTE abundance of {\LyudaStar} and the scaled solar $r$-abundance amounts to 0.17~dex and 0.24~dex for the data of Arlandini1999 and Bisterzo2014, respectively. Our calculations show that NLTE leads to a 0.15~dex lower Ba, but 0.10~dex higher Eu abundance of {\LyudaStar} and, thus, removes the difference in Ba abundance between {\LyudaStar} and SSr of Bisterzo2014, when assuming the NLTE corrections for Gd-Tm abundances to be similar to those for Eu.

\section{Thorium in $r$-process enhanced stars}\label{Sect:Th}

It would be rather challenging to determine the stellar age of {\LyudaStar} by comparing the observed
Th-to-stable neutron-capture element-abundance ratios with 
the corresponding initial values at the time when the star was born, log(Th/$X$)$_0$:
\begin{displaymath}
\tau = 46.7\,\mathrm{Gyr}\,\left[\log({\rm Th}/X)_0 - \log({\rm Th}/X)_{obs}\right].
\end{displaymath}

\noindent For {\LyudaStar}, the ratio log(Th/Eu)$_{obs} = -0.33$ (LTE) or $-0.35$ (NLTE) deviates by only a little, 0.01 or 0.03~dex, from the
solar $r$-residual ratio log(Th/Eu)$_0 = -0.32$ (Bisterzo2014), resulting in too low an age of $\tau < 1.5$~Gyr. Since the Sun is approximately 4.5\,Gyr old, a corresponding correction of +0.1~dex accounting for the thorium radioactive decay was introduced to the solar current thorium abundance. 
To be as old as the galactic halo VMP stars, {\LyudaStar} should have an approximately 0.3~dex lower Th abundance. Figure~\ref{Fig:th_line} displays the effect of a variation in Th abundance on the
    synthetic spectrum of \ion{Th}{ii} 4019\,{\AA} and \ion{Th}{ii} 3741\,{\AA}. It is evident that the stellar Th abundance cannot be significantly lower than $\eps{Th} = -1.63$.
The observational error represented by the dispersion in the measurements of
multiple lines around the mean amounts to $\sigma_{\eps{}} =$ 0.02~dex for Th/Eu of {\LyudaStar}. The systematic errors linked to our choice of stellar parameters have a common sign and very similar magnitudes for lines of \ion{Eu}{ii} and \ion{Th}{ii}, i.e. 0.08~dex and 0.09~dex, respectively. Thus, the total uncertainty in Th/Eu is small, and it is translated to an uncertainty of less than 1.5~Gyr for the age of {\LyudaStar}.  

Application of the above formula to the Th/$X$ pairs involving other $r$-process elements leads to $\tau = 1.9$ to 3.3~Gyr for Gd, Ho, Er, and Os, $\tau = 7.0$ and 7.9~Gyr for Dy and Ir, and negative age for Tb and Tm.
It is clear that Th does appear to be enhanced in {\LyudaStar} to a higher level than observed for elements of the second and third $r$-process peaks, and {\LyudaStar} belongs to the group of the so-called actinide boost stars.

\begin{figure} 
  \centering
  \resizebox{88mm}{!}{\includegraphics{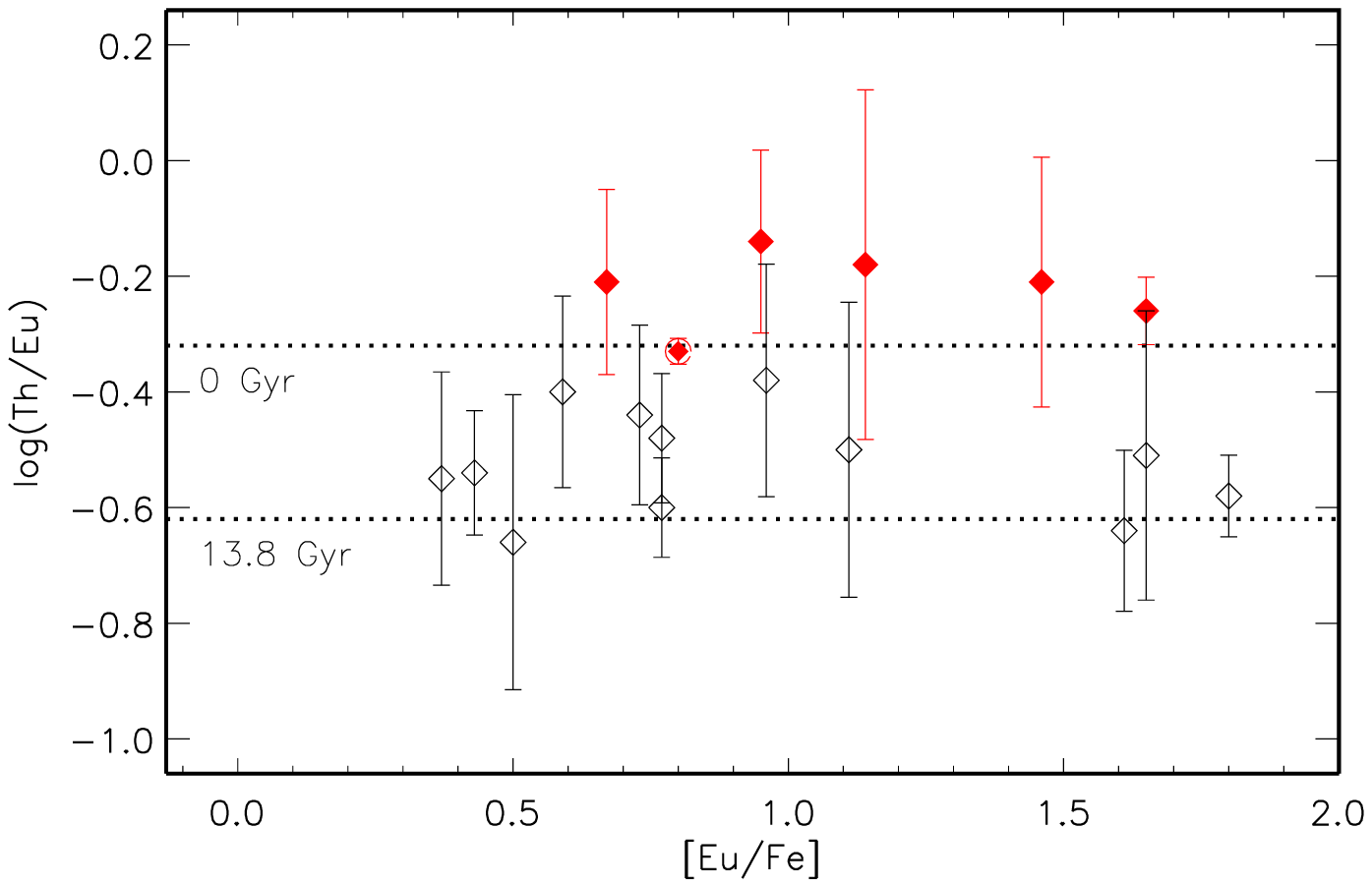}}
  \caption{\label{Fig:th_eu} Th/Eu abundance ratios of the \citet{Roederer2009} sample, {\LyudmilasStar} (Paper~V), CS\,30315-029 \citep{2014arXiv1404.0234S}, and {\LyudaStar} (filled rhomb inside open circle). Stars with high Th/Eu ratios (log(Th/Eu) $\ge -0.35$) are shown by filled rhombi and the remaining stars with open rhombi. The horizontal lines indicate the ratios expected if a sample of material had a given age, assuming the solar $r$-process of Bisterzo2014.}
\end{figure}

The first discovered star with an anomalously high Th/Eu abundance ratio of log(Th/Eu) $= -0.22$ was CS\,31082-001 \citep{hill2002}. \citet{Roederer2009} revised Eu and Th abundances of 15 MP stars that reveal pure $r$-process neutron-capture elements. 
Figure\,\ref{Fig:th_eu} displays the Th/Eu abundance ratios of the \citet{Roederer2009} sample and also {\LyudmilasStar} (Paper~V), CS\,30315-029 \citep{2014arXiv1404.0234S}, and {\LyudaStar}. 
It appears that 12 of the 18 stars have no obvious
enhancement of thorium with respect to the scaled solar $r$-process pattern, and for each of them, its stellar age agrees within the error bars with the cosmic age $13.772\pm0.059$~Gyr \citep{2013ApJS..208...20B} derived from the updated results of the {\em Wilkinson Microwave Anisotropy Probe (WMAP)}. For example, $\tau = 14.9\pm6.5$~Gyr was calculated for CS\,22892-052 with log(Th/Eu) $= -0.64$. 
The remaining six stars, namely, CS\,31082-001, CS\,30306-132 \citep[discovered by][]{Honda2004}, CS\,31078-018 \citep{Lai2008}, HE\,1219-0312 \citep{HE1219}, CS\,30315-029 \citep{2014arXiv1404.0234S}, and {\LyudaStar}, exhibit an actinide boost, and their ages cannot be derived
when only a single radioactive element Th is detected. Indeed, a negative age of $\tau = -4.7$~Gyr is calculated for CS\,31082-001, when using the Th/Eu abundance ratio. This implies that thorium in the actinide boost stars was overproduced compared with the normal Th/Eu stars and the SS matter. Only the detection of the second actinide, uranium, made possible an estimation of the stellar age of CS\,31082-001 through analysis of U/Th; i.e., $\tau = 12.5\pm3$~Gyr was first obtained by \citet{2001Natur.409..691C}, and the revised value is $\tau = 14.0\pm2.4$~Gyr \citep{hill2002}. This provided solid evidence that different $r$-process nucleosynthesis events may produce significantly different yields in the actinide region ($Z \ge 90$). 
To find out whether variations in progenitor mass or explosion energy, or other intrinsic and
environmental factors, or all of these, influence the $r$-process yields for the heaviest elements, more measurements of Th and U abundances in stars should be done.
 
\section{Conclusions}\label{Sect:DiscussionConclusions}

We revised the stellar parameters and performed a detailed abundance analysis of the VMP giant {\LyudaStar} using high-quality VLT/UVES spectra and refined theoretical methods of line-formation modelling. The effective temperature, $\teffm = 4710$~K, previously derived in Paper~II was confirmed through analysis of the H$\alpha$ and H$\beta$ line wings in {\LyudaStar}. The surface gravity, log~g = 1.65, the iron abundance, [Fe/H] $= -2.63$, and the microturbulence velocity, $\xi_t = 1.7$\,\kms, were calculated from the NLTE ionisation balance between \ion{Fe}{i} and \ion{Fe}{ii}.

Accurate abundances for a total of 38 elements from C to Th were determined in {\LyudaStar}. For each chemical species, the dispersion in the single line measurements around the mean does not exceed 0.12~dex. The investigated star was found to be deficient in carbon, as expected for a giant star with $\teffm < 4800$~K. The elements in the range from Na to Zn reveal a typical behaviour of the galactic halo VMP stars. The Na--Zn abundance pattern of {\LyudaStar} is well fitted by the yields of a single supernova of 14.4 $M_\odot$, similar to the \citet{Lai2008} stellar sample. 

We confirmed that {\LyudaStar} is $r$-process enhanced, having $[r/\mathrm{Fe}] = 0.80\pm0.06$.
The investigated star and four benchmark r-II stars;
i.e., CS\,22892-052 \citep{Sneden2003}, CS\,31082-001 \citep{2013A&A...550A.122S}, HE\,1219-0312
\citep{HE1219}, and HE\,1523-091 \citep{Sneden2008} have very similar abundance patterns of the elements in the range from Sr to Ir. Hence, neutron-capture elements beyond Sr and up to Ir in {\LyudaStar} have a common origin in the classical main $r$-process. 
Applying the third criterion, [Sr/Eu] $< -0.8$ \citep{HE2327}, in addition to the two, [Eu/Fe] $> +1$ and [Ba/Eu] $< 0$, as suggested by \citet{HERESI}, makes membership to r-II stars have a physical sense related to an origin of the neutron-capture elements in the star; i.e., an r-II star can be defined as a star having neutron-capture elements originating in a single $r$-process.

We tested the solar $r$-process pattern based on recent $s$-process calculations of \citet{2014arXiv1403.1764B} and found that the elements in the range from
 Ba to Ir in {\LyudaStar} match it very well. No firm
  conclusion can be drawn about the relationship between the first
  neutron-capture peak elements, Sr to Ru, in {\LyudaStar} and the solar
  $r$-process, owing to the uncertainty in the solar $r$-process.

The star {\LyudaStar} has an anomalously high Th/Eu abundance ratio, so that radioactive age dating  results in a stellar age of $\tau = 1.5\pm1.5$~Gyr that is not expected for a very metal-poor halo star. This is the sixth star in the group of actinide boost stars. Understanding the mechanisms resulting in different yields in the actinide region from different $r$-process events is a challenge for nucleosynthesis theory and requires studying larger samples of r-I and r-II stars.

\begin{acknowledgements}
  
This work was supported by Sonderforschungsbereich SFB
 881 ``The Milky Way System'' (subprojects A4 and A5) of 
the German   Research Foundation (DFG).
L.M. is supported by the Russian Foundation for Basic Research (grant 14-02-91153) and the Swiss National Science Foundation (SCOPES project No.~IZ73Z0-152485).
We made use the NIST and VALD databases.

\end{acknowledgements}

\bibliography{mashonkina,atomic_data,nlte,mp_stars,references}
\bibliographystyle{aa}

\Online

\longtab{3}{
\begin{longtable}[]{rcccrlcrrrrcl}   
\caption{\label{linelist} Line data and abundances from an analysis of {\LyudaStar}. $\Gamma_6$ corresponds to 10\,000~K. Column 6 gives references to the adopted $gf$-values. Column 13 gives references to the sources of the used IS and HFS data and adopted $\Gamma_6$-values.}  \\
\hline \\ 
 Z & Atom/ & $\lambda$ & $E_{exc}$ & $\log gf$ & \multicolumn{1}{c}{Ref} & \multicolumn{3}{c}{$\eps{}$} & [X/Fe] & $W_\lambda$ &  $\log \Gamma_6/N_{\rm H}$ & Note, ref. \\
\cline{7-9}
   &  mol  & (\AA)     & (eV)     &   &   & Solar & LTE  & NLTE & & (m\AA) & (rad/s$\cdot$cm$^3$)& \\
\hline 
 1 & 2 & 3 & 4 & \multicolumn{1}{c}{5} & \multicolumn{1}{c}{6} & 7 & \multicolumn{1}{c}{8} & \multicolumn{1}{c}{9} & \multicolumn{1}{c}{10} & \multicolumn{1}{c}{11} & 12 & \multicolumn{1}{c}{13} \\
\hline \\
\endfirsthead
\caption{continued.}\\
\hline \\
 Z & Atom/ & $\lambda$ & $E_{exc}$ & $\log gf$ & Ref & \multicolumn{3}{c}{$\eps{}$} & [X/Fe] & $W_\lambda$ &  $\log \Gamma_6/N_{\rm H}$ & Note, ref. \\
\cline{7-9}
   &  mol  & (\AA)     & (eV)     &   &   & Solar & LTE  & NLTE & & (m\AA) & (rad/s$\cdot$cm$^3$) & \\
\hline 
 1 & 2 & 3 & 4 & \multicolumn{1}{c}{5} & \multicolumn{1}{c}{6} & 7 & \multicolumn{1}{c}{8} & \multicolumn{1}{c}{9} & \multicolumn{1}{c}{10} & \multicolumn{1}{c}{11} & 12 & \multicolumn{1}{c}{13} \\
\hline  \\   
\endhead
\hline
\endfoot
\hline
\endlastfoot
  3 & Li I & 6707.80 & 0.00 &  0.167 & NIST8 &  1.10 & $\le -$0.1 & &  & Syn & $-$7.574 & IS:SRE95 \\
    &      &         &      &        &       &      &       &      &      &      &        & $\Gamma_6$:ABO \\
\multicolumn{13}{c}{ } \\
  6 & CH  &  4310-4312.5  &   & & BCB05 &  8.39 &  5.13 &        & $-$0.63 & Syn & & \\
  6 & CH  &  4314	  &   &   & BCB05 &  8.39 &  5.16 &        & $-$0.60 & Syn & & \\
  6 & CH  &  4362-4364.5  &   & & BCB05 &  8.39 &  5.14 &        & $-$0.62 & Syn & & \\
  6 & CH  &  4366-4367.0  &   & & BCB05 &  8.39 &  5.16 &        & $-$0.60 & Syn & & \\
 \multicolumn{13}{c}{ } \\
  7 & NH  &  3416.64 &   & $-$0.94 & K94& 7.83 &  4.91 &      & $-$0.29 & Syn & & \\
 \multicolumn{13}{c}{ } \\
 11 & Na I & 5889.96 & 0.00 &    0.11 & NIST8 & 6.30 &  3.57 & 3.13 & $-$0.54 & Syn & $-$7.67 & $\Gamma_6$:ABO \\
 11 & Na I & 5895.93 & 0.00 & $-$0.19 & NIST8 & 6.30 &  3.57 & 3.13 & $-$0.54 & Syn & $-$7.67 & $\Gamma_6$:ABO \\
 \multicolumn{13}{c}{ } \\
 12 & Mg I & 4571.10 & 0.00 & $-$5.62 & NIST8 & 7.54 &  4.98 & 5.22 &    0.31 & Syn & $-$7.77 & $\Gamma_6$:ABO \\
 12 & Mg I & 4702.99 & 4.33 & $-$0.44 & CT90  & 7.54 &  4.98 & 5.01 &    0.10 & Syn & $-$6.99 & $\Gamma_6$:M13 \\
 12 & Mg I & 5528.41 & 4.33 & $-$0.50 & CT90  & 7.54 &  5.02 & 5.03 &    0.12 & Syn & $-$7.18 & $\Gamma_6$:M13 \\
 \multicolumn{13}{c}{ } \\
 13 & Al I & 3961.52 & 0.01 & $-$0.34 & NIST8 & 6.47 &  2.85 & 2.99 & $-$0.85 & Syn & $-$7.32 & $\Gamma_6$:ABO \\
 \multicolumn{13}{c}{ } \\
 14 & Si I & 3905.53 & 1.91 & $-$1.04 & BL91  & 7.52 &  4.99 & 4.97 &  0.08   & Syn & $-$7.36 & $\Gamma_6$:ABO \\
  \multicolumn{13}{c}{ } \\
 20 & Ca I & 4425.44 & 1.88 & $-$0.36 & SON75 & 6.33 &  3.60 & 3.86 &  0.16   & 36.8 & $-$7.432 & $\Gamma_6$:MKP07 \\
 20 & Ca I & 5349.46 & 2.71 & $-$0.31 & SR81  &  6.33 &  3.74 & 4.05 &  0.35 & 10.1 & $-$7.652 & $\Gamma_6$:S81 \\
 20 & Ca I & 5588.75 & 2.53 &    0.36 & SR81  &  6.33 &  3.62 & 3.92 &  0.22 & 39.5 & $-$7.628 & $\Gamma_6$:S81 \\
 20 & Ca I & 5857.45 & 2.93 &    0.24 & SR81  &  6.33 &  3.69 & 3.87 &  0.17 & 17.0 & $-$7.316 & $\Gamma_6$:S81 \\
 20 & Ca I & 6122.22 & 1.89 & $-$0.31 & SON75 & 6.33 &  3.73 & 3.96 &  0.26 & 53.4 & $-$7.189 & $\Gamma_6$:ABO \\
 20 & Ca I & 6162.17 & 1.90 & $-$0.09 & SON75 & 6.33 &  3.70 & 3.90 &  0.20 & 64.5 & $-$7.189 & $\Gamma_6$:ABO \\
 20 & Ca I & 6169.56 & 2.53 & $-$0.48 & SR81  &  6.33 &  3.68 & 3.97 &  0.27 & 10.5 & $-$7.264 & $\Gamma_6$:S81 \\
 20 & Ca I & 6439.08 & 2.53 &    0.39 & SR81  &  6.33 &  3.65 & 3.83 &  0.13 & 45.6 & $-$7.704 & $\Gamma_6$:S81 \\
 20 & Ca I & 6449.81 & 2.52 & $-$0.50 & SR81  &  6.33 &  3.69 & 3.93 &  0.23 & 10.6 & $-$7.652 & $\Gamma_6$:S81 \\
 20 & Ca I & 6493.78 & 2.52 & $-$0.11 & SR81  &  6.33 &  3.70 & 3.95 &  0.25 & 23.3 & $-$7.704 & $\Gamma_6$:S81 \\
\multicolumn{13}{c}{ } \\
 21 & Sc II & 3567.70 & 0.00 & $-$0.48 & LD89 & 3.07 &  0.39 &    & $-$0.05 & Syn & & HFS:MPS95 \\
 21 & Sc II & 4314.09 & 0.62 & $-$0.10 & LD89 & 3.07 &  0.31 &    & $-$0.13 & Syn & & the same \\
 21 & Sc II & 4400.39 & 0.60 & $-$0.54 & LD89 & 3.07 &  0.26 &    & $-$0.18 & Syn & & the same \\
 21 & Sc II & 4415.56 & 0.60 & $-$0.67 & LD89 & 3.07 &  0.31 &    & $-$0.13 & Syn & & the same \\
 \multicolumn{13}{c}{ } \\
 22 & Ti I & 3998.64 & 0.05 &   0.02 & LGW13 &  4.90 &  2.21 &    &  0.20 &  67.6 & $-$7.654 & $\Gamma_6$:ABO \\
 22 & Ti I & 4533.25 & 0.85 &   0.54 & LGW13 &  4.90 &  2.22 &    &  0.21 &  50.2 & $-$7.593 & the same \\
 22 & Ti I & 4534.78 & 0.84 &   0.35 & LGW13 &  4.90 &  2.19 &    &  0.18 &  39.1 & $-$7.596 & the same \\
 22 & Ti I & 4981.73 & 0.85 &   0.57 & LGW13 &  4.90 &  2.28 &    &  0.27 &  58.2 & $-$7.626 & the same \\
 22 & Ti I & 4991.06 & 0.84 &   0.45 & LGW13 &  4.90 &  2.29 &    &  0.28 &  53.0 & $-$7.629 & the same \\
 22 & Ti I & 5022.87 & 0.83 & $-$0.33 & LGW13 & 4.90 &  2.22 &    &  0.21 &  13.8 & $-$7.633 & the same \\
 22 & Ti I & 5024.85 & 0.82 & $-$0.53 & LGW13 & 4.90 &  2.38 &    &  0.37 &  13.1 & $-$7.635 & the same \\
 22 & Ti I & 5039.96 & 0.02 & $-$1.08 & LGW13 & 4.90 &  2.31 &    &  0.30 &  25.7 & $-$7.720 & the same \\
 22 & Ti I & 5064.66 & 0.05 & $-$0.94 & LGW13 & 4.90 &  2.28 &    &  0.27 &  29.2 & $-$7.719 & the same \\
 22 & Ti I & 5210.39 & 0.05 & $-$0.82 & LGW13 & 4.90 &  2.27 &    &  0.26 &  35.5 & $-$7.724 & the same \\
 22 & Ti II & 4028.34 & 1.89 & $-$0.92 & WL13 & 4.90 &  2.62 &    &  0.36 &  56.8 & & \\
 22 & Ti II & 4053.83 & 1.89 & $-$1.07 & WL13 & 4.90 &  2.44 &    &  0.17 &   39.3 & & \\
 22 & Ti II & 4394.05 & 1.22 & $-$1.77 & WL13 & 4.90 &  2.65 &    &  0.38 &   59.6 & & \\
 22 & Ti II & 4395.85 & 1.24 & $-$1.93 & WL13 & 4.90 &  2.62 &    &  0.35 &   48.0 & & \\
 22 & Ti II & 4399.77 & 1.24 & $-$1.20 & WL13 & 4.90 &  2.74 &    &  0.47 &   89.1 & & \\
 22 & Ti II & 4417.72 & 1.16 & $-$1.19 & PTP01 & 4.90 &  2.76 &   &  0.49 &   94.0 & & \\
 22 & Ti II & 4418.33 & 1.24 & $-$1.99 & WL13  & 4.90 &  2.64 &    &  0.37 &   45.5 & & \\
 22 & Ti II & 4444.56 & 1.12 & $-$2.20 & WL13  & 4.90 &  2.71 &    &  0.44 &   46.1 & & \\
 22 & Ti II & 4450.48 & 1.08 & $-$1.52 & WL13  & 4.90 &  2.73 &    &  0.46 &   83.7 & & \\
 22 & Ti II & 4464.45 & 1.16 & $-$1.81 & PTP01 & 4.90 &  2.72 &   &  0.45 &   65.1 & & \\
 22 & Ti II & 4470.86 & 1.16 & $-$2.02 & PTP01 & 4.90 &  2.55 &   &  0.28 &   45.0 & & \\
 22 & Ti II & 5185.91 & 1.89 & $-$1.41 & WL13  & 4.90 &  2.54 &    &  0.27 &   33.2 & & \\
 22 & Ti II & 5188.70 & 1.58 & $-$1.05 & PTP01 & 4.90 &  2.70 &   &  0.43 &   82.4 & & \\
 22 & Ti II & 5226.55 & 1.57 & $-$1.26 & PTP01 & 4.90 &  2.61 &   &  0.34 &   67.8 & & \\
 22 & Ti II & 5336.79 & 1.58 & $-$1.60 & WL13  & 4.90 &  2.60 &    &  0.33 &   47.2 & & \\
 22 & Ti II & 5381.01 & 1.57 & $-$1.97 & WL13  & 4.90 &  2.60 &    &  0.33 &   28.6 & & \\
 \multicolumn{13}{c}{ } \\
 23 & V  II & 3530.76 & 1.07 & $-$0.47 & BGF89 & 4.00 &  1.09 &   & $-$0.28 & Syn & & \\
 23 & V  II & 3545.19 & 1.10 & $-$0.26 & BGF89 & 4.00 &  1.14 &   & $-$0.23 & Syn & & \\
 23 & V  II & 3592.03 & 1.10 & $-$0.26 & BGF89 & 4.00 &  1.14 &   & $-$0.23 & Syn & & \\
 \multicolumn{13}{c}{ } \\
 24 & Cr I & 5206.04 & 0.94 &   0.02   & SLS07 & 5.64 &  2.62 &   & $-$0.13 & 83.6 & $-$7.597 & $\Gamma_6$:ABO \\
 24 & Cr I & 5296.69 & 0.98 &  $-$1.36 & SLS07 & 5.64 &  2.73 &   & $-$0.02 & 17.4 & $-$7.621 & the same \\
 24 & Cr I & 5345.81 & 1.00 &  $-$0.95 & SLS07 & 5.64 &  2.70 &   & $-$0.05 & 32.8 & $-$7.620 & the same \\
 24 & Cr I & 5348.33 & 1.00 &  $-$1.21 & SLS07 & 5.64 &  2.60 &   & $-$0.15 & 17.2 & $-$7.620 & the same \\
 24 & Cr I & 5409.80 & 1.03 &  $-$0.67 & SLS07 & 5.64 &  2.64 &   & $-$0.11 & 42.8 & $-$7.620 & the same \\
 24 & Cr II & 4558.65 & 4.07 & $-$0.45 & PGB93 & 5.65 &  3.10 &   &  0.08 & Syn & & \\
 24 & Cr II & 4588.20 & 4.07 & $-$0.63 & PGB93 & 5.65 &  3.08 &   &  0.06 & Syn & & \\
 24 & Cr II & 4634.07 & 4.07 & $-$0.98 & NLL06 & 5.65 &  2.99 &   & $-$0.03 & Syn & & \\
 24 & Cr II & 4848.23 & 3.86 & $-$1.00 & NLL06 & 5.65 &  3.01 &   & $-$0.01 & Syn & & \\
 \multicolumn{13}{c}{ } \\
 25 & Mn I & 4033.06 & 0.00 & -0.64 & DLS11 &  5.37 &  2.05 & & -0.43 & Syn &  -7.738 & HFS:LGB03 \\
    &      &         &      &        &       &       &       & &       &     &         & $\Gamma_6$:ABO \\
 25 & Mn I & 4041.35 & 2.11 &  0.28  & DLS11 &  5.37 &  2.35 & & $-$0.13 & Syn &  -7.701 & the same \\
 25 & Mn I & 4055.54 & 2.13 & $-$0.08 & DLS11 &  5.37 &  2.44 & & $-$0.04 & Syn &  -7.698 & the same \\
 25 & Mn I & 4823.52 & 2.32 & 0.12    & DLS11 &  5.37 &  2.44 & & $-$0.04 & Syn &  -7.600 & the same \\
 25 & Mn II & 3460.31 & 1.81 & $-$0.64 & KG00 & 5.37 &  2.63 &    & $-$0.11 & Syn & & HFS:HSR99 \\
 25 & Mn II & 3482.90 & 1.83 & $-$0.84 & KG00 & 5.37 &  2.63 &    & $-$0.11 & Syn & & the same \\
 25 & Mn II & 3488.67 & 1.85 & $-$0.95 & KG00 & 5.37 &  2.60 &    & $-$0.14 & Syn & & the same \\
 \multicolumn{13}{c}{ } \\
 26 & Fe I & 3917.18 & 0.99 &  $-$2.15 & OWL91 & 7.45 &  4.82 & 5.04 &  0.22 &   91.7 & -7.695 & $\Gamma_6$:AB \\
 26 & Fe I & 3949.95 & 2.18 &  $-$1.25 & OWL91 & 7.45 &  4.60 & 4.84 &  0.02 &   61.8 & -7.820 & the same  \\ 
 26 & Fe I & 4132.90 & 2.85 &  $-$1.01 & OWL91 & 7.45 &  4.57 & 4.82 &  0.00 &   34.9 & -7.659 & the same  \\ 
 26 & Fe I & 4147.67 & 1.49 &  $-$2.07 & OWL91 & 7.45 &  4.70 & 4.91 &  0.09 &   69.5 & -7.648 & the same  \\ 
 26 & Fe I & 4216.18 & 0.00 &  $-$3.36 & FMW88 & 7.45 &  4.86 & 5.12 &  0.30 &   98.8 & -7.797 & the same  \\ 
 26 & Fe I & 4260.47 & 2.40 &   0.08   & OWL91 & 7.45 &  4.53 & 4.70 & $-$0.12 &110.0 & -7.274 & the same  \\ 
 26 & Fe I & 4445.47 & 0.09 &  $-$5.44 & BIP79 & 7.45 &  4.79 & 5.00 &  0.18 &    7.5 & -7.816 & the same  \\ 
 26 & Fe I & 4602.94 & 1.49 &  $-$2.21 & OWL91 & 7.45 &  4.65 & 4.89 &  0.07 &   65.6 & -7.790 & the same  \\ 
 26 & Fe I & 4647.43 & 2.95 &  $-$1.35 & OWL91 & 7.45 &  4.67 & 4.93 &  0.11 &   21.0 & -7.685 & the same  \\ 
 26 & Fe I & 4920.50 & 2.83 &   0.07   & OWL91 & 7.45 &  4.56 & 4.75 & $-$0.07 & 94.8 & -7.276 & the same  \\ 
 26 & Fe I & 4966.09 & 3.33 &  $-$0.87 & OWL91 & 7.45 &  4.67 & 4.97 &  0.15 &   23.2 & -7.218 & the same  \\ 
 26 & Fe I & 4994.13 & 0.92 &  $-$2.97 & OWL91 & 7.45 &  4.67 & 4.89 &  0.07 &   66.6 & -7.744 & the same  \\ 
 26 & Fe I & 5001.86 & 3.88 &   0.01   & FMW88 & 7.45 &  4.43 & 4.73 & $-$0.09 & 22.9 & -7.273 & the same  \\ 
 26 & Fe I & 5014.94 & 3.94 &  $-$0.30 & OWL91 & 7.45 &  4.65 & 4.93 &  0.11 &   16.8 & -7.268 & the same  \\ 
 26 & Fe I & 5051.63 & 0.92 &  $-$2.77 & OWL91 & 7.45 &  4.82 & 5.03 &  0.21 &   84.1 & -7.746 & the same  \\ 
 26 & Fe I & 5068.77 & 2.94 &  $-$1.04 & OWL91 & 7.45 &  4.55 & 4.81 & $-$0.01 & 31.1 & -7.265 & the same  \\ 
 26 & Fe I & 5150.84 & 0.99 &  $-$3.04 & OWL91 & 7.45 &  4.62 & 4.83 &  0.01 &   55.7 & -7.742 & the same  \\ 
 26 & Fe I & 5166.28 & 0.00 &  $-$4.20 & OWL91 & 7.45 &  4.81 & 5.05 &  0.23 &   70.8 & -7.826 & the same  \\ 
 26 & Fe I & 5192.34 & 3.00 &  $-$0.42 & OWL91 & 7.45 &  4.47 & 4.74 & $-$0.08 & 57.5 & -7.266 & the same  \\ 
 26 & Fe I & 5194.94 & 1.56 &  $-$2.02 & OWL91 & 7.45 &  4.66 & 4.86 &  0.04 &   75.4 & -7.680 & the same  \\ 
 26 & Fe I & 5198.71 & 2.22 &  $-$2.14 & BPSS  & 7.45 &  4.65 & 4.87 &  0.05 &   25.2 & -7.600 & the same  \\ 
 26 & Fe I & 5216.27 & 1.61 &  $-$2.08 & OWL91 & 7.45 &  4.62 & 4.81 & $-$0.01 & 67.0 & -7.674 & the same  \\ 
 26 & Fe I & 5217.39 & 3.21 &  $-$1.07 & BKK91 & 7.45 &  4.51 & 4.75 & $-$0.07 & 15.9 & -7.220 & the same \\ 
 26 & Fe I & 5232.94 & 2.94 &  $-$0.06 & OWL91 & 7.45 &  4.51 & 4.74 & $-$0.08 & 82.1 & -7.288 & the same \\ 
 26 & Fe I & 5247.06 & 0.09 &  $-$4.95 & BIP79 & 7.45 &  4.75 & 4.98 &  0.16 &   22.0 & -7.826 & the same \\ 
 26 & Fe I & 5250.21 & 0.12 &  $-$4.94 & BIP79 & 7.45 &  4.76 & 4.98 &  0.16 &   21.0 & -7.832 & the same \\ 
 26 & Fe I & 5281.79 & 3.04 &  $-$0.83 & OWL91 & 7.45 &  4.51 & 4.78 & $-$0.04 & 35.0 & -7.266 & the same  \\ 
 26 & Fe I & 5283.62 & 3.24 &  $-$0.52 & OWL91 & 7.45 &  4.61 & 4.87 &  0.05 &   43.7 & -7.221 & the same  \\ 
 26 & Fe I & 5324.18 & 3.21 &  $-$0.10 & BKK91 & 7.45 &  4.43 & 4.69 & $-$0.13 & 59.3 & -7.235 & the same  \\ 
 26 & Fe I & 5339.93 & 3.27 &  $-$0.72 & OWL91 & 7.45 &  4.58 & 4.84 &  0.02 &   30.4 & -7.221 & the same  \\ 
 26 & Fe I & 5364.87 & 4.45 &   0.23  & OWL91 &  7.45 &  4.65 & 4.97 &  0.15 &   15.7 & -7.136 & the same  \\ 
 26 & Fe I & 5367.47 & 4.41 &   0.44  & OWL91 &  7.45 &  4.50 & 4.84 &  0.02 &   18.8 & -7.153 & the same  \\ 
 26 & Fe I & 5369.96 & 4.37 &   0.54  & OWL91 &  7.45 &  4.44 & 4.79 & $-$0.03 & 22.1 & -7.179 & the same  \\ 
 26 & Fe I & 5383.37 & 4.31 &   0.64  & OWL91 &  7.45 &  4.45 & 4.81 & $-$0.01 & 30.2 & -7.219 & the same  \\ 
 26 & Fe I & 5393.17 & 3.24 & $-$0.72 & BKK91 &  7.45 &  4.53 & 4.79 & $-$0.03 & 29.8 & -7.235 & the same  \\ 
 26 & Fe I & 5410.91 & 4.47 &   0.40  & OWL91 &  7.45 &  4.43 & 4.77 & $-$0.05 & 13.5 & -7.132 & the same  \\ 
 26 & Fe I & 5415.20 & 4.39 &   0.64  & OWL91 &  7.45 &  4.42 & 4.80 & $-$0.02 & 25.0 & -7.182 & the same  \\
 26 & Fe I & 5434.52 & 1.01 &  $-$2.13 & OWL91 & 7.45 &  4.79 & 4.94 &  0.12 &  110.0 & -7.749 & the same  \\
 26 & Fe I & 5445.04 & 4.39 &  $-$0.02 & FMW88 & 7.45 &  4.74 & 5.06 &  0.24 &   13.2 & -7.189 & the same  \\
 26 & Fe I & 5506.78 & 0.99 &  $-$2.80 & OWL91 & 7.45 &  4.79 & 4.98 &  0.16 &   80.3 & -7.753 & the same  \\
 26 & Fe I & 5576.09 & 3.43 &  $-$1.00 & FMW88 & 7.45 &  4.70 & 4.94 &  0.12 &   16.5 & -7.200 & the same \\
 26 & Fe I & 5586.76 & 3.37 &  $-$0.10 & BKK91 & 7.45 &  4.43 & 4.68 & $-$0.13 & 49.9 & -7.221 & the same  \\
 26 & Fe I & 5615.64 & 3.33 &   0.05   & BKK91 & 7.45 &  4.39 & 4.64 & $-$0.18 & 59.0 & -7.234 & the same  \\
 26 & Fe I & 6065.48 & 2.61 &  $-$1.53 & BPS82 & 7.45 &  4.62 & 4.83 &  0.01 &   33.8 & -7.636 & the same \\
 26 & Fe I & 6136.62 & 2.45 &  $-$1.41 & OWL91 & 7.45 &  4.65 & 4.86 &  0.04 &   53.0 & -7.609 & the same  \\
 26 & Fe I & 6137.69 & 2.59 &  $-$1.35 & OWL91 & 7.45 &  4.60 & 4.81 & $-$0.01 & 43.9 & -7.589 & the same  \\
 26 & Fe I & 6191.56 & 2.43 &  $-$1.42 & OWL91 & 7.45 &  4.57 & 4.79 & $-$0.03 & 50.1 & -7.615 & the same  \\
 26 & Fe I & 6213.43 & 2.22 &  $-$2.48 & OWL91 & 7.45 &  4.59 & 4.81 & $-$0.01 & 13.6 & -7.704 & the same \\
 26 & Fe I & 6252.56 & 2.40 &  $-$1.69 & BPSS  & 7.45 &  4.66 & 4.88 &  0.06 &   41.3 & -7.680 & the same \\
 26 & Fe I & 6393.60 & 2.43 &  $-$1.43 & OWL91 & 7.45 &  4.50 & 4.75 & $-$0.07 & 46.0 & -7.622 & the same  \\
 26 & Fe I & 6400.00 & 3.60 &  $-$0.29 & BKK91 & 7.45 &  4.48 & 4.77 & $-$0.05 & 30.2 & -7.232 & the same  \\
 26 & Fe I & 6411.65 & 3.65 &  $-$0.59 & BKK91 & 7.45 &  4.52 & 4.79 & $-$0.03 & 16.9 & -7.224 & the same \\
 26 & Fe I & 6421.35 & 2.28 &  $-$2.03 & BPSS  & 7.45 &  4.66 & 4.86 &  0.04 &   31.9 & -7.792 & the same \\
 26 & Fe I & 6430.85 & 2.18 &  $-$1.95 & OWL91 & 7.45 &  4.59 & 4.81 & $-$0.01 & 39.2 & -7.704 & the same \\
 26 & Fe II & 4491.40 & 2.86 &  $-$2.71 & MB09 & 7.45 &  4.80 & 4.80 & $-$0.02 & 25.1 & -7.880 & the same \\
 26 & Fe II & 4508.29 & 2.86 &  $-$2.44 & MB09 & 7.45 &  4.88 & 4.89 &  0.07 &   42.4 & -7.870 & the same \\
 26 & Fe II & 4582.83 & 2.84 &  $-$3.18 & MB09 & 7.45 &  4.73 & 4.73 & $-$0.09 &  9.5 & -7.880 & the same \\
 26 & Fe II & 4620.52 & 2.83 &  $-$3.21 & MB09 & 7.45 &  4.69 & 4.70 & $-$0.12 &  8.6 & -7.880 & the same \\
 26 & Fe II & 4923.93 & 2.89 &  $-$1.26 & MB09 & 7.45 &  4.85 & 4.82 &  0.01 &   98.5 & -7.890 & the same  \\
 26 & Fe II & 5018.44 & 2.89 &  $-$1.10 & MB09 & 7.45 &  4.89 & 4.86 &  0.04 &  109.0 & -7.890 & the same  \\
 26 & Fe II & 5197.58 & 3.23 &  $-$2.22 & MB09 & 7.45 &  4.84 & 4.85 &  0.03 &   31.4 & -7.880 & the same  \\
 26 & Fe II & 5234.62 & 3.22 &  $-$2.18 & MB09 & 7.45 &  4.82 & 4.83 &  0.01 &   33.3 & -7.880 & the same \\ 
 26 & Fe II & 5264.81 & 3.23 &  $-$3.13 & MB09 & 7.45 &  4.91 & 4.92 &  0.10 &    6.4 & -7.880 & the same \\ 
 26 & Fe II & 5284.11 & 2.89 &  $-$3.11 & MB09 & 7.45 &  4.85 & 4.85 &  0.03 &   13.5 & -7.890 & the same \\ 
 26 & Fe II & 6247.56 & 3.89 &  $-$2.30 & MB09 & 7.45 &  4.72 & 4.74 & $-$0.08 &  5.3 & -7.880 & the same \\
 26 & Fe II & 6456.38 & 3.90 &  $-$2.05 & MB09 & 7.45 &  4.85 & 4.87 &  0.05 &   11.6 & -7.880 & the same \\
 \multicolumn{13}{c}{ } \\
 27 & Co I  & 3412.33 & 0.51 &     0.03 & CSS82 & 4.92 &  1.91 &     & $-$0.12 & Syn & -7.666 & HFS:P96 \\
    &      &         &      &        &       &       &       & &     &     &         & $\Gamma_6$:ABO \\
 27 & Co I  & 3489.40 & 0.92 &     0.15 & CSS82 & 4.92 &  2.00 &     & $-$0.03 & Syn & -7.610 & the same \\
 27 & Co I  & 4121.31 & 0.92 &  $-$0.30 & NKW99 & 4.92 &  2.08 &     &  0.05   & Syn & -7.724 & the same\\
 27 & Co II & 3501.72 & 2.20 & $-$1.00  & SLW85 & 4.92 &  2.16 &     & $-$0.13 & Syn & & \\
 \multicolumn{13}{c}{ } \\
 28 & Ni I  & 3413.47 & 0.16 & $-$1.48 & FMW88  & 6.23 &  3.15 & & $-$0.19 & Syn & -7.690 & $\Gamma_6$:ABO \\
 28 & Ni I  & 3413.93 & 0.11 & $-$1.72 & FMW88 & 6.23 &  3.25 & & $-$0.09 & Syn & -7.785 & the same \\
 28 & Ni I  & 3783.52 & 0.42 & $-$1.31 & FMW88  & 6.23 &  3.34 & &  0.00 & Syn & -7.780 & the same \\
 28 & Ni I  & 3807.14 & 0.42 & $-$1.18 & FMW88 & 6.23 &  3.45 & &  0.11 & Syn & -7.694 & the same \\
 28 & Ni I  & 3858.29 & 0.42 & $-$0.97 & FMW88 & 6.23 &  3.35 & &  0.01 & Syn & -7.700 & the same \\
 28 & Ni I  & 5035.37 & 3.63 &  0.29   & WL97a & 6.23 &  3.51 & &  0.17   & Syn & -7.231 & the same \\
 28 & Ni II & 3769.46 & 3.10 & $-$1.72 & K03 & 6.23 &  3.75 &    &  0.15 & Syn & & \\
 \multicolumn{13}{c}{ } \\
 30 & Zn I & 4722.16 & 4.01 & $-$0.37 & RL12 & 4.62 &  2.43 & 2.41 &  0.42 & Syn & & \\
 30 & Zn I & 4810.54 & 4.08 & $-$0.15 & RL12 & 4.62 &  2.44 & 2.42 &  0.43 & Syn & & \\
 \multicolumn{13}{c}{ } \\
 38 & Sr II & 4077.72 & 0.00 &  0.15 & RCW80 & 2.92 &  0.16 & 0.23 & $-$0.06 & Syn & -7.792 & HFS:BBH83 \\
    &      &         &      &        &       &       &       & &       &     &         & $\Gamma_6$:MZG08 \\
 38 & Sr II & 4215.54 & 0.00 &$-$0.17& RCW80 & 2.92 &  0.16 & 0.16 & $-$0.13 & Syn & -7.792 & HFS:BBH83 \\
    &      &         &      &        &       &       &       & &       &     &         & $\Gamma_6$:MZG08 \\
 \multicolumn{13}{c}{ } \\
 39 & Y  II & 3549.01 & 0.13 & $-$0.28 & HLG82 &  2.21 & $-$0.36 &        & 0.06 & Syn & & \\
 39 & Y  II & 3600.74 & 0.18 &  0.28 & HLG82 &  2.21 & $-$0.56 &        & $-$0.14 & Syn & & \\
 39 & Y  II & 3611.04 & 0.13 &  0.01 & HLG82 &  2.21 & $-$0.77 &        & $-$0.35 & Syn & & \\
 39 & Y  II & 3774.33 & 0.13 &  0.21 & HLG82 &  2.21 & $-$0.56 &        & $-$0.14 & Syn & & \\
 39 & Y  II & 3788.69 & 0.10 & $-$0.07 & HLG82 &  2.21 & $-$0.59 &        & $-$0.17 & Syn & & \\
 39 & Y  II & 3950.35 & 0.10 & $-$0.49 & HLG82 &  2.21 & $-$0.55 &        & $-$0.13 & Syn & & \\
 39 & Y  II & 4398.01 & 0.13 & $-$1.00 & HLG82 &  2.21 & $-$0.54 &        & $-$0.12 & Syn & & \\
 39 & Y  II & 4883.68 & 1.08 &  0.07   & HLG82 &  2.21 & $-$0.62 &        & $-$0.20 & Syn & & \\
 39 & Y  II & 5087.43 & 1.08 & $-$0.17 & HLG82 &  2.21 & $-$0.64 &        & $-$0.22 & Syn & & \\
 \multicolumn{13}{c}{ } \\
 40 & Zr II & 3404.83 & 0.36 & $-$0.49 & LNA06 &  2.58 & $-$0.06 &        & $-$0.01 & Syn & & \\
 40 & Zr II & 3457.56 & 0.56 & $-$0.47 & MBM06 &  2.58 & $-$0.11 &        & $-$0.06 & Syn & & \\
 40 & Zr II & 3479.39 & 0.71 &  0.18   & LNA06 &  2.58 &  0.06 &        &  0.11 & Syn & & \\
 40 & Zr II & 3481.17 & 0.80 &  0.16   & CC83  &  2.58 &  0.07 &        &  0.12 & Syn & & \\
 40 & Zr II & 3499.57 & 0.41 & $-$1.06 & LNA06 &  2.58 &  0.04 &        &  0.09 & Syn & & \\
 40 & Zr II & 3505.67 & 0.16 & $-$0.39 & LNA06 &  2.58 &  0.09 &        &  0.14 & Syn & & \\
 40 & Zr II & 3551.95 & 0.10 & $-$0.36 & LNA06 &  2.58 &  0.07 &        &  0.12 & Syn & & \\
 40 & Zr II & 3766.82 & 0.41 & $-$0.83 & LNA06 &  2.58 &  0.17 &        &  0.22 & Syn & & \\
 40 & Zr II & 3991.13 & 0.76 & $-$0.31 & LNA06 &  2.58 &  0.06 &        &  0.11 & Syn & & \\
 40 & Zr II & 3998.97 & 0.56 & $-$0.52 & LNA06 &  2.58 &  0.12 &        &  0.17 & Syn & & \\
 40 & Zr II & 4161.21 & 0.71 & $-$0.59 & LNA06 &  2.58 &  0.09 &        &  0.14 & Syn & & \\
 40 & Zr II & 4208.98 & 0.71 & $-$0.51 & LNA06 &  2.58 &  0.12 &        &  0.17 & Syn & & \\
 \multicolumn{13}{c}{ } \\
 42 & Mo I & 3864.11 & 0.00 & $-$0.01 & FW96 &  1.92 & $-$0.84 &        &  0.13 & Syn & & \\
 \multicolumn{13}{c}{ } \\
 44 & Ru I & 3728.02 & 0.00 &  0.27 & WSL94 &  1.84 & $-$0.45 &        &  0.60 & Syn & & \\
 \multicolumn{13}{c}{ } \\
 56 & Ba II & 5853.67 & 0.60 & $-$1.00 & RCW80 &  2.17 & $-$0.25 &  $-$0.27 &  0.19 & Syn & -7.584 & $\Gamma_6$:ABO \\
 56 & Ba II & 6141.71 & 0.70 & $-$0.08 & RCW80 &  2.17 & $-$0.18 &  $-$0.35 &  0.11 & Syn & -7.584 & the same \\
 56 & Ba II & 6496.90 & 0.60 & $-$0.38 & RCW80 &  2.17 & $-$0.10 &  $-$0.33 &  0.13 & Syn & -7.584 & the same \\
 \multicolumn{13}{c}{ } \\
 57 & La II & 3849.01 & 0.00 & $-$0.45 & LBS01 &  1.14 & $-$1.12 &        &  0.37 & Syn & & HFS:LBS01 \\
 57 & La II & 3949.10 & 0.40 &  0.49 & LBS01 &  1.14 & $-$1.16 &        &  0.33 & Syn & & the same \\
 57 & La II & 3988.51 & 0.40 &  0.21 & LBS01 &  1.14 & $-$1.15 &        &  0.34 & Syn & & the same \\
 57 & La II & 3995.74 & 0.17 & $-$0.06 & LBS01 &  1.14 & $-$1.11 &        &  0.38 & Syn & & the same \\
 57 & La II & 4077.34 & 0.24 & $-$0.06 & LBS01 &  1.14 & $-$1.01 &        &  0.48 & Syn & & the same \\
 57 & La II & 4086.71 & 0.00 & $-$0.07 & LBS01 &  1.14 & $-$1.14 &        &  0.35 & Syn & & the same \\
 57 & La II & 4196.55 & 0.32 & $-$0.30 & LBS01 &  1.14 & $-$1.09 &        &  0.40 & Syn & & the same \\
 57 & La II & 4920.98 & 0.13 & $-$0.58 & LBS01 &  1.14 & $-$0.99 &        &  0.50 & Syn & & the same \\
 \multicolumn{13}{c}{ } \\
 58 & Ce II & 3942.15 & 0.00 & $-$0.22 & LSC09 &  1.61 & $-$0.75 &        &  0.27 & Syn & & \\
 58 & Ce II & 3992.38 & 0.45 & $-$0.22 & LSC09 &  1.61 & $-$0.75 &        &  0.27 & Syn & & \\
 58 & Ce II & 3999.24 & 0.30 &  0.06   & LSC09 &  1.61 & $-$0.75 &        &  0.27 & Syn & & \\
 58 & Ce II & 4073.47 & 0.48 &  0.21   & LSC09 &  1.61 & $-$0.72 &        &  0.30 & Syn & & \\
 58 & Ce II & 4083.22 & 0.70 &  0.27   & LSC09 &  1.61 & $-$0.73 &        &  0.29 & Syn & & \\
 58 & Ce II & 4127.36 & 0.68 &  0.31   & LSC09 &  1.61 & $-$0.77 &        &  0.25 & Syn & & \\
 58 & Ce II & 4137.64 & 0.52 &  0.40   & LSC09 &  1.61 & $-$0.74 &        &  0.28 & Syn & & \\
 58 & Ce II & 4222.60 & 0.12 & $-$0.15 & LSC09 &  1.61 & $-$0.68 &        &  0.34 & Syn & & \\
 58 & Ce II & 4486.91 & 0.30 & $-$0.18 & LSC09 &  1.61 & $-$0.73 &        &  0.29 & Syn & & \\
 58 & Ce II & 4562.36 & 0.48 &  0.21   & LSC09 &  1.61 & $-$0.68 &        &  0.34 & Syn & & \\
 \multicolumn{13}{c}{ } \\
 59 & Pr II & 4179.40 & 0.20 &  0.48 & ILW01 &  0.76 & $-$1.23 &        &  0.64 & Syn & & HFS:G89 \\
 59 & Pr II & 4408.82 & 0.00 &  0.18 & ILW01 &  0.76 & $-$1.29 &        &  0.58 & Syn & & the same \\
 59 & Pr II & 4222.93 & 0.05 &  0.27 & ILW01 &  0.76 & $-$1.25 &        &  0.62 & Syn & & the same \\
 \multicolumn{13}{c}{ } \\
 60 & Nd II & 3784.25 & 0.38 &  0.15   & DLS03 &  1.45 & $-$0.72 &        &  0.46 & Syn & & \\
 60 & Nd II & 3838.98 & 0.00 & $-$0.24 & DLS03 &  1.45 & $-$0.68 &        &  0.50 & Syn & & \\
 60 & Nd II & 3991.74 & 0.00 & $-$0.26 & DLS03 &  1.45 & $-$0.58 &        &  0.60 & Syn & & \\
 60 & Nd II & 4018.82 & 0.06 & $-$0.85 & DLS03 &  1.45 & $-$0.66 &        &  0.52 & Syn & & \\
 60 & Nd II & 4021.33 & 0.32 & $-$0.10 & DLS03 &  1.45 & $-$0.69 &        &  0.49 & Syn & & \\
 60 & Nd II & 4023.00 & 0.56 &  0.04   & DLS03 &  1.45 & $-$0.64 &        &  0.54 & Syn & & \\
 60 & Nd II & 4059.95 & 0.20 & $-$0.52 & DLS03 &  1.45 & $-$0.71 &        &  0.47 & Syn & & \\
 60 & Nd II & 4061.08 & 0.47 &  0.55   & DLS03 &  1.45 & $-$0.61 &        &  0.57 & Syn & & \\
 60 & Nd II & 4069.26 & 0.06 & $-$0.57 & DLS03 &  1.45 & $-$0.56 &        &  0.62 & Syn & & \\
 60 & Nd II & 4135.32 & 0.63 & $-$0.07 & DLS03 &  1.45 & $-$0.60 &        &  0.58 & Syn & & \\
 60 & Nd II & 4156.08 & 0.18 &  0.16   & DLS03 &  1.45 & $-$0.56 &        &  0.62 & Syn & & \\
 60 & Nd II & 4177.33 & 0.06 & $-$0.10 & DLS03 &  1.45 & $-$0.62 &        &  0.56 & Syn & & \\
 60 & Nd II & 4232.37 & 0.06 & $-$0.47 & DLS03 &  1.45 & $-$0.64 &        &  0.54 & Syn & & \\
 60 & Nd II & 4385.66 & 0.20 & $-$0.30 & DLS03 &  1.45 & $-$0.64 &        &  0.54 & Syn & & \\
 60 & Nd II & 4446.38 & 0.20 & $-$0.35 & DLS03 &  1.45 & $-$0.61 &        &  0.57 & Syn & & \\
 60 & Nd II & 4462.98 & 0.56 &  0.04   & DLS03 &  1.45 & $-$0.56 &        &  0.62 & Syn & & \\
 60 & Nd II & 4563.22 & 0.18 & $-$0.88 & DLS03 &  1.45 & $-$0.44 &        &  0.74 & Syn & & \\
 60 & Nd II & 4706.54 & 0.00 & $-$0.71 & DLS03 &  1.45 & $-$0.58 &        &  0.60 & Syn & & \\
 60 & Nd II & 5319.82 & 0.55 & $-$0.14 & DLS03 &  1.45 & $-$0.62 &        &  0.56 & Syn & & \\
 \multicolumn{13}{c}{ } \\
 62 & Sm II & 3993.31 & 0.04 & $-$0.93 & LDS06 &  1.00 & $-$0.92 &        &  0.71 & Syn & & \\
 62 & Sm II & 4318.94 & 0.28 & $-$0.25 & LDS06 &  1.00 & $-$0.89 &        &  0.74 & Syn & & \\
 62 & Sm II & 4434.32 & 0.38 & $-$0.07 & LDS06 &  1.00 & $-$0.89 &        &  0.74 & Syn & & \\
 62 & Sm II & 4467.34 & 0.66 &  0.15   & LDS06 &  1.00 & $-$0.91 &        &  0.72 & Syn & & \\
 62 & Sm II & 4519.63 & 0.54 & $-$0.35 & LDS06 &  1.00 & $-$0.89 &        &  0.74 & Syn & & \\
 \multicolumn{13}{c}{ } \\
 63 & Eu II & 3819.67 & 0.00 &  0.51 & LWD01 &  0.52 & $-$1.31 &  $-$1.24 &  0.87 & Syn & & HFS:LWD01 \\
 63 & Eu II & 3907.11 & 0.21 &  0.17 & LWD01 &  0.52 & $-$1.31 &  $-$1.16 &  0.95 & Syn & & the same \\
 63 & Eu II & 4129.72 & 0.00 &  0.22 & LWD01 &  0.52 & $-$1.29 &  $-$1.20 &  0.91 & Syn & & the same \\
 63 & Eu II & 4205.02 & 0.00 &  0.21 & LWD01 &  0.52 & $-$1.28 &  $-$1.22 &  0.89 & Syn & & the same \\
 \multicolumn{13}{c}{ } \\
 64 & Gd II & 3768.40 & 0.08 &  0.21 & DLS06 &  1.11 & $-$0.74 &        &  0.78 & Syn & & \\
 64 & Gd II & 3796.38 & 0.03 &  0.02 & DLS06 &  1.11 & $-$0.71 &        &  0.81 & Syn & & \\
 64 & Gd II & 4130.37 & 0.73 &  0.14 & DLS06 &  1.11 & $-$0.84 &        &  0.68 & Syn & & \\
 64 & Gd II & 4215.02 & 0.43 & $-$0.44 & DLS06 &  1.11 & $-$0.79 &        &  0.73 & Syn & & \\
 \multicolumn{13}{c}{ } \\
 65 & Tb II & 3568.51 & 0.00 &  0.36 & LWC01 &  0.28 & $-$1.56 &        &  0.79 & Syn & & HFS:LWB01 \\
 65 & Tb II & 3702.85 & 0.13 &  0.44 & LWC01 &  0.28 & $-$1.56 &        &  0.79 & Syn & & the same \\
 65 & Tb II & 3848.73 & 0.00 &  0.28 & LWC01 &  0.28 & $-$1.56 &        &  0.79 & Syn & & the same \\
 \multicolumn{13}{c}{ } \\
 66 & Dy II & 3407.80 & 0.00 &  0.18   & WLN00 &  1.13 & $-$0.50 &        &  1.00 & Syn & & \\
 66 & Dy II & 3445.57 & 0.00 & $-$0.15 & WLN00 &  1.13 & $-$0.57 &        &  0.93 & Syn & & \\
 66 & Dy II & 3454.32 & 0.10 & $-$0.14 & WLN00 &  1.13 & $-$0.77 &        &  0.73 & Syn & & \\
 66 & Dy II & 3460.97 & 0.00 & $-$0.07 & WLN00 &  1.13 & $-$0.57 &        &  0.93 & Syn & & \\
 66 & Dy II & 3506.81 & 0.10 & $-$0.60 & WLN00 &  1.13 & $-$0.62 &        &  0.88 & Syn & & \\
 66 & Dy II & 3531.71 & 0.00 &  0.77   & WLN00 &  1.13 & $-$0.52 &        &  0.98 & Syn & & \\
 66 & Dy II & 3536.02 & 0.54 &  0.53   & WLN00 &  1.13 & $-$0.67 &        &  0.83 & Syn & & \\
 66 & Dy II & 3538.52 & 0.00 & $-$0.02 & WLN00 &  1.13 & $-$0.67 &        &  0.83 & Syn & & \\
 66 & Dy II & 3550.22 & 0.59 &  0.27   & WLN00 &  1.13 & $-$0.67 &        &  0.83 & Syn & & \\
 66 & Dy II & 3563.15 & 0.10 & $-$0.36 & WLN00 &  1.13 & $-$0.47 &        &  1.03 & Syn & & \\
 66 & Dy II & 3694.81 & 0.10 & $-$0.11 & WLN00 &  1.13 & $-$0.65 &        &  0.85 & Syn & & \\
 66 & Dy II & 3944.68 & 0.00 &  0.11   & WLN00 &  1.13 & $-$0.47 &        &  1.03 & Syn & & \\
 66 & Dy II & 3996.69 & 0.59 & $-$0.26 & WLN00 &  1.13 & $-$0.57 &        &  0.93 & Syn & & \\
 66 & Dy II & 4077.96 & 0.10 & $-$0.04 & WLN00 &  1.13 & $-$0.52 &        &  0.98 & Syn & & \\
 \multicolumn{13}{c}{ } \\
 67 & Ho II & 3416.44 & 0.08 &  0.26 & LSC04 &  0.51 & $-$1.24 &        &  0.88 & Syn & & HFS:LSC04 \\
 67 & Ho II & 3456.00 & 0.00 &  0.76 & LSC04 &  0.51 & $-$1.40 &        &  0.72 & Syn & & the same \\
 67 & Ho II & 3796.67 & 0.00 &  0.16 & LSC04 &  0.51 & $-$1.23 &        &  0.89 & Syn & & the same \\
 67 & Ho II & 3810.60 & 0.00 &  0.19 & LSC04 &  0.51 & $-$1.30 &        &  0.82 & Syn & & the same \\
 \multicolumn{13}{c}{ } \\
 68 & Er II & 3499.10 & 0.06 &  0.29   & LSC08 &  0.96 & $-$0.93 &        &  0.74 & Syn & & \\
 68 & Er II & 3633.54 & 0.00 & $-$0.53 & LSC08 &  0.96 & $-$1.08 &        &  0.59 & Syn & & \\
 68 & Er II & 3692.65 & 0.06 &  0.28   & LSC08 &  0.96 & $-$0.83 &        &  0.84 & Syn & & \\
 68 & Er II & 3729.52 & 0.00 & $-$0.59 & LSC08 &  0.96 & $-$0.83 &        &  0.84 & Syn & & \\
 68 & Er II & 3786.84 & 0.00 & $-$0.52 & LSC08 &  0.96 & $-$0.78 &        &  0.89 & Syn & & \\
 68 & Er II & 3830.48 & 0.00 & $-$0.22 & LSC08 &  0.96 & $-$0.88 &        &  0.79 & Syn & & \\
 68 & Er II & 3896.23 & 0.06 & $-$0.12 & LSC08 &  0.96 & $-$0.86 &        &  0.81 & Syn & & \\
 68 & Er II & 3938.63 & 0.00 & $-$0.52 & MCS75 &  0.96 & $-$0.98 &        &  0.69 & Syn & & \\
 \multicolumn{13}{c}{ } \\
 69 & Tm II & 3462.20 & 0.00 &  0.03   & WL97b &  0.14 & $-$1.81 &        &  0.68 & Syn & & \\
 69 & Tm II & 3700.26 & 0.03 & $-$0.38 & WL97b &  0.14 & $-$1.71 &        &  0.78 & Syn & & \\
 69 & Tm II & 3701.36 & 0.00 & $-$0.54 & WL97b &  0.14 & $-$1.81 &        &  0.68 & Syn & & \\
 69 & Tm II & 3795.76 & 0.03 & $-$0.23 & WL97b &  0.14 & $-$1.74 &        &  0.75 & Syn & & \\
 69 & Tm II & 3848.02 & 0.00 & $-$0.14 & WL97b &  0.14 & $-$1.74 &        &  0.75 & Syn & & \\
 \multicolumn{13}{c}{ } \\
 70 & Yb II & 3694.19 & 0.00 & $-$0.30 & BDM98 &  0.86 & $-$0.91 &        &  0.86 & Syn & & HFS:MGH94 \\
 \multicolumn{13}{c}{ } \\
 76 & Os I  & 4260.85 & 0.00 & $-$1.43 & IAN03 &  1.45 & $-$0.44 &        &  1.00 & Syn & & \\
 \multicolumn{13}{c}{ } \\
 77 & Ir I & 3513.630 & 0.00 & $-$1.21 & IAN03 &  1.38 & $-$0.25 &        &  1.26 & Syn & & HFS:CSB05 \\
 77 & Ir I & 3800.120 & 0.00 & $-$1.44 & IAN03 &  1.38 & $-$0.39 &        &  1.12 & Syn & & HFS:CSB05 \\
 \multicolumn{13}{c}{ } \\
 82 & Pb I & 4057.81 & 1.32 & $-$0.17 & BGP00 &  2.00 & $\le -$0.78 &  $\le -$0.37 & $\le$0.26 & Syn & & \\
 \multicolumn{13}{c}{ } \\
 90 & Th II & 3741.18 & 0.19 & $-$0.17 & NZL02 &  0.08 & $-$1.65 &  $-$1.55 &  1.00 & Syn & & \\
 90 & Th II & 4019.13 & 0.00 & $-$0.23 & NZL02 &  0.08 & $-$1.62 &  $-$1.56 &  0.99 & Syn & & \\
\hline 
\end{longtable}
\begin{table}
\begin{tabular}{ll}
\multicolumn{1}{l}{ABO} & \multicolumn{1}{l}{\citet{1995MNRAS.276..859A,1997MNRAS.290..102B,1998MNRAS.296.1057B};} \\
\multicolumn{1}{l}{ } & \multicolumn{1}{l}{\citet{1998MNRAS.300..863B,2005A&A...435..373B}; Barklem (2012, private comm.);} \\
\multicolumn{1}{l}{BBH83} & \multicolumn{1}{l}{\citet{1983HyInt..15..177B};} \\
\multicolumn{1}{l}{BCB05} & \multicolumn{1}{l}{\citet{HERESII};} \\
\multicolumn{1}{l}{BDM98} & \multicolumn{1}{l}{\citet{Biemont_Yb};} \\
\multicolumn{1}{l}{BGF89} & \multicolumn{1}{l}{\citet{1989A&A...209..391B};} \\
\multicolumn{1}{l}{BGP00} & \multicolumn{1}{l}{\citet{pb_gf_biemont};} \\
\multicolumn{1}{l}{BIP79} & \multicolumn{1}{l}{\citet{1979MNRAS.186..633B};} \\
\multicolumn{1}{l}{BKK91} & \multicolumn{1}{l}{\citet{1991A&A...248..315B};}\\
\multicolumn{1}{l}{BL91} & \multicolumn{1}{l}{ \citet{1991PhRvA..44.7134O};}\\
\multicolumn{1}{l}{BPSS} & \multicolumn{1}{l}{\citet{1982MNRAS.199...43B};} \\
\multicolumn{1}{l}{BPS82} & \multicolumn{1}{l}{\citet{1982MNRAS.201..595B};} \\
\multicolumn{1}{l}{CC83} & \multicolumn{1}{l}{ \citet{1983MNRAS.203..651C};}\\
\multicolumn{1}{l}{CSB05} & \multicolumn{1}{l}{ \citet{2005ApJ...627..238C};}\\
\multicolumn{1}{l}{CSS82} & \multicolumn{1}{l}{ \citet{1982ApJ...260..395C};}\\
\multicolumn{1}{l}{CT90} & \multicolumn{1}{l}{ \citet{1990JQSRT..43..207C};}\\
\multicolumn{1}{l}{DLS03} & \multicolumn{1}{l}{ \citet{Hartog_Nd};}\\
\multicolumn{1}{l}{DLS06} & \multicolumn{1}{l}{ \citet{Hartog_Gd};}\\
\multicolumn{1}{l}{DLS11} & \multicolumn{1}{l}{ \citet{DLSSC};}\\
\multicolumn{1}{l}{FMW88} & \multicolumn{1}{l}{ \citet{1988JPCRD..17S....F};}\\
\multicolumn{1}{l}{FW96} & \multicolumn{1}{l}{ \citet{FW96};}\\
\multicolumn{1}{l}{G89} & \multicolumn{1}{l}{ \citet{1989PhyS...39..694G};}\\
\multicolumn{1}{l}{HLG82} & \multicolumn{1}{l}{\citet{1982ApJ...261..736H};}\\
\multicolumn{1}{l}{HSR99} & \multicolumn{1}{l}{\citet{1999MNRAS.306..107H};}\\
\multicolumn{1}{l}{IAN03} & \multicolumn{1}{l}{\citet{IAN};}\\
\multicolumn{1}{l}{ILW01} & \multicolumn{1}{l}{\citet{Ivarsson_Pr};}\\
\multicolumn{1}{l}{K94} & \multicolumn{1}{l}{ \citet{Kur94};}\\
\multicolumn{1}{l}{K03} & \multicolumn{1}{l}{ \citet{K03};}\\
\multicolumn{1}{l}{KG00} & \multicolumn{1}{l}{  \citet{2000ApJ...531.1173K};}\\
\multicolumn{1}{l}{LBS01} & \multicolumn{1}{l}{\citet{2001ApJ...556..452L};}\\
\multicolumn{1}{l}{LD89} & \multicolumn{1}{l}{ \citet{1989JOSAB...6.1457L};}\\
\multicolumn{1}{l}{LDS06} & \multicolumn{1}{l}{ \citet{Lawler_Sm};}\\
\multicolumn{1}{l}{LGB03} & \multicolumn{1}{l}{ \citet{2003A&A...404.1153L};}\\
\multicolumn{1}{l}{LGW13} & \multicolumn{1}{l}{ \citet{Lawler2013_ti1};}\\
\multicolumn{1}{l}{LNA06} & \multicolumn{1}{l}{ \citet{LNAJ};}\\
\multicolumn{1}{l}{LSC04} & \multicolumn{1}{l}{ \citet{Lawler_Ho};}\\
\multicolumn{1}{l}{LSC08} & \multicolumn{1}{l}{ \citet{Lawler_Er};}\\
\multicolumn{1}{l}{LSC09} & \multicolumn{1}{l}{ \citet{Lawler_Ce};}\\
\multicolumn{1}{l}{LWB01} & \multicolumn{1}{l}{ \citet{Lawler_Tb_hfs};}\\
\multicolumn{1}{l}{LWC01} & \multicolumn{1}{l}{ \citet{Lawler_Tb};}\\ 
\multicolumn{1}{l}{LWD01} & \multicolumn{1}{l}{ \citet{Lawler_Eu};}\\
\multicolumn{1}{l}{M13} & \multicolumn{1}{l}{ \citet{mash_mg13};}\\
\multicolumn{1}{l}{MB09} & \multicolumn{1}{l}{ \citet{MB09};}\\ 
\multicolumn{1}{l}{MBM06} & \multicolumn{1}{l}{ \citet{2006MNRAS.367..754M};}\\ 
\multicolumn{1}{l}{MCS75} & \multicolumn{1}{l}{ \citet{MCS75};}\\ 
\multicolumn{1}{l}{MGH94} & \multicolumn{1}{l}{ \citet{hfs-yb2};}\\
\multicolumn{1}{l}{MKP07} & \multicolumn{1}{l}{ \citet{mash_ca};}\\
\multicolumn{1}{l}{MPS95} & \multicolumn{1}{l}{ \citet{1995AJ....109.2757M};}\\
\multicolumn{1}{l}{MZG08} & \multicolumn{1}{l}{ \citet{Mashonkina2008};}\\
\multicolumn{1}{l}{NIST8} & \multicolumn{1}{l}{ \citet{NIST08};}\\ 
\multicolumn{1}{l}{NKW99} & \multicolumn{1}{l}{ \citet{1999ApJS..122..557N};}\\ 
\multicolumn{1}{l}{NLL06} & \multicolumn{1}{l}{ \citet{NLLN};}\\ 
\multicolumn{1}{l}{NZL02} & \multicolumn{1}{l}{ \citet{Nilsson_Th};}\\
\multicolumn{1}{l}{OWL91} & \multicolumn{1}{l}{ \citet{1991PhRvA..44.7134O};}\\ 
\multicolumn{1}{l}{P96} & \multicolumn{1}{l}{ \citet{hfs-co1};}\\
\multicolumn{1}{l}{PGB93} & \multicolumn{1}{l}{ \citet{PGBH};}\\
\multicolumn{1}{l}{PTP01} & \multicolumn{1}{l}{ \citet{2001ApJS..132..403P};}\\ 
\multicolumn{1}{l}{RCW80} & \multicolumn{1}{l}{ \citet{1980wtpa.book.....R};}\\ 
\multicolumn{1}{l}{RL12} & \multicolumn{1}{l}{ \citet{2012ApJ...750...76R};}\\
\multicolumn{1}{l}{S81} & \multicolumn{1}{l}{ \citet{1981A&A...103..351S};}\\
\multicolumn{1}{l}{SLS07} & \multicolumn{1}{l}{ \citet{SLS07};}\\
\multicolumn{1}{l}{SLW85} & \multicolumn{1}{l}{ \citet{co2_gf};}\\
\multicolumn{1}{l}{SR81} & \multicolumn{1}{l}{ \citet{1981JPhB...14.4015S};}\\
\multicolumn{1}{l}{SON75} & \multicolumn{1}{l}{\citet{1975A&A....38....1S};}\\ 
\multicolumn{1}{l}{SRE95} & \multicolumn{1}{l}{\citet{1995PhRvA..52.2682S};}\\
\multicolumn{1}{l}{WB88} & \multicolumn{1}{l}{ \citet{WB88};}\\
\multicolumn{1}{l}{WL13} & \multicolumn{1}{l}{ \citet{2013_gf_ti2};}\\
\multicolumn{1}{l}{WL97a} & \multicolumn{1}{l}{ \citet{1997ApJS..110..163W};}\\
\multicolumn{1}{l}{WL97b} & \multicolumn{1}{l}{ \citet{Wickliffe_Tm};}\\
\multicolumn{1}{l}{WLN00} & \multicolumn{1}{l}{ \citet{Wickliffe_Dy};}\\
\multicolumn{1}{l}{WSL94} & \multicolumn{1}{l}{ \citet{WSL94};}\\
\end{tabular}
\end{table}
}

\end{document}